\documentclass[runningheads,a4paper]{llncs}
\usepackage{etex}
\reserveinserts{28}
\pdfoutput=1

\usepackage[linesnumbered,lined,boxed,commentsnumbered,ruled]{algorithm2e}

\SetAlFnt{\small}
\SetAlCapFnt{\small}
\SetAlCapNameFnt{\small}
\SetAlCapHSkip{0pt}
\IncMargin{-\parindent}
\usepackage{ textcomp }
\usepackage{ comment }
\usepackage[color=yellow!60,textsize=footnotesize,obeyDraft,draft]{todonotes}
\usepackage{colortbl}
\usepackage{booktabs}
\usepackage{capt-of}
\usepackage{makecell}
\usepackage{textcomp }
\usepackage{multirow}
\usepackage{listings}
\usepackage{graphicx}
\usepackage{colortbl}
\usepackage{booktabs}
\usepackage{amsmath}
\usepackage[font=footnotesize,labelfont=bf]{subcaption}
\usepackage[hidelinks]{hyperref}
\usepackage[capitalize]{cleveref}
\usepackage{tikz}
\usepackage{pgfplots}
\usepackage[para,online,flushleft]{threeparttable}
\usepackage{multirow}
\usepackage{wasysym}
\usepackage{amssymb}
\usepackage{enumitem}
\usetikzlibrary{arrows}
\usepackage{subcaption}
\usepackage{breakcites}
\usepackage{url}
\newcounter{mnotei}
\newcolumntype{L}[1]{>{\raggedright\let\newline\\\arraybackslash\hspace{0pt}}m{#1}}
\newcolumntype{C}[1]{>{\centering\let\newline\\\arraybackslash\hspace{0pt}}m{#1}}
\newcolumntype{R}[1]{>{\raggedleft\let\newline\\\arraybackslash\hspace{0pt}}m{#1}}

\newcommand{\includegraphicsmaybe}[2]{
    \IfFileExists{#2}{\includegraphics[#1]{#2}}{
    \detokenize{File #2 is missing, maybe you need to run plots.py?}
}}

\begin{document}
\mainmatter

\title{Lost in translation: Exposing hidden compiler optimization opportunities}

\titlerunning{Lost in translation: Exposing hidden compiler optimization opportunities}

\author{Kyriakos Georgiou, Zbigniew Chamski, Andres \nolinebreak[4] Amaya \nolinebreak[4] Garcia, David May, Kerstin Eder}

\authorrunning{K. Georgiou et al.}
\institute{University of Bristol, UK}
\tocauthor{Authors' Instructions}
\maketitle

\makeatletter
\renewcommand\subsubsection{\@startsection{subsubsection}{3}{\z@}%
                       {-18\p@ \@plus -4\p@ \@minus -4\p@}%
                       {4\p@ \@plus 2\p@ \@minus 2\p@}%
                       {\normalfont\normalsize\bfseries\boldmath
                        \rightskip=\z@ \@plus 8em\pretolerance=10000 }}
\makeatother

\begin{abstract}
Existing iterative compilation and machine-learning-based optimization techniques have been proven very successful in achieving better optimizations than the standard optimization levels of a compiler. However, they were not engineered to support the tuning of a compiler's optimizer as part of the compiler's daily development cycle. In this paper, we first establish the required properties which a technique must exhibit to enable such tuning. We then introduce an enhancement to the classic nightly routine testing of compilers which exhibits all the required properties, and thus, is capable of driving the improvement and tuning of the compiler's common optimizer. This is achieved by leveraging resource usage and compilation information collected while systematically exploiting prefixes of the transformations applied at standard optimization levels. Experimental evaluation using the LLVM v6.0.1 compiler demonstrated that the new approach was able to reveal hidden cross-architecture and architecture-dependent potential optimizations on two popular processors: the Intel i5-6300U and the Arm Cortex-A53-based Broadcom BCM2837 used in the Raspberry Pi 3B+. As a case study, we demonstrate how the insights from our approach enabled us to identify and remove a significant shortcoming of the CFG simplification pass of the LLVM v6.0.1 compiler.
\end{abstract}

\section{Introduction}
\label{sec:introduction}

Compilers are at the heart of software development. Their primary goal is to increase software productivity. They are a key software engineering tool, automating the translation from high-level languages to machine code. This sets a major challenge to compiler engineers as they have to support a vast number of architectures and programming languages and adapt to their rapid advances. To mitigate this, modern compilers, such as the LLVM~\cite{LattnerLLVM2004,LLVM} and GCC~\cite{GCC_compiler} compilers, are designed to be modular. For example, they make use of a common, target-independent optimizer across all the architectures and programming languages supported. The common optimizer~\cite{Lattner:2012}, operates on a target-independent Intermediate Representation (IR). Each supported programming language is translated to this IR through a customized version of the compiler's front-end, and then the common optimizer can perform the generic optimizations on the generated IR. Finally, the IR can be translated to the machine code of a supported architecture through the target-specific back-end. Thus, all the supported architectures and programming languages can benefit from the generic optimizations implemented in the common optimizer. In this paper, we also refer to the LLVM's IR optimizer as the ``common optimizer''.

The common optimizer exposes to the software developer a large number of available code optimizations via compiler flags. The LLVM optimizer has 58 transformation flags~\cite{LLVM:passes2020}, where a code transformation can apply an optimization, or it can facilitate an IR structure which enables the application of other optimizations. In the more mature GCC compiler, as of version 9.3.0 approximately 110 out of the 229 optimization flags~\cite{GCC_opt_flags} control the passes operating on the target-independent IR. The challenge then becomes to select and order the flags to create optimization configurations that can achieve the best resource usage possible for a given program and architecture. Due to the huge number of possible flag combinations and their possible orderings, it is impractical to explore the complete space of optimization configurations. Furthermore, many of these optimizations are tunable, with their behavior dependent on a number of heuristics and quantitative parameters. The selection of the actual heuristic and the determination of the numerical values of parameters are in many cases hidden from the compiler user, but are exposed to the compiler developers and can be partially influenced by the selection of an overall optimization level, e.g.\ in the case of the LLVM compiler toolchain. Thus, finding optimal optimization configurations is still an open challenge.

To address this, compilers offer standard optimization levels, typically -O0, -O1, -O2, -O3 and -Os which are predefined sequences of optimizations. These sequences are tuned through each new compiler release to perform well on a number of micro-benchmarks and a range of mainstream architectures. Starting from the -O0 level, which has no optimizations enabled, and moving to level -O3, each level offers more aggressive optimizations with the main focus being performance, while -Os is focused on optimizing code size. New standard optimization levels are often introduced into modern compilers to address more specific optimization needs and constraints, such as the -Oz optimization level offered by the LLVM compiler, which tries to reduce code size even further than -Os, and the -Og optimization level introduced by GCC, which aims at improving the debugging experience of optimized code.
Still, such optimizations are demonstrably not optimal, as iterative compilation and machine-learning-based (MLB) approaches can find optimization sequences that offer better resource usage than the standard optimization levels on a particular program and architecture~\cite{Boyle:2018}. The main idea of such approaches is to find good optimization sequences by exploiting only a fraction of the optimization space~\cite{Ashouri:survey}. 

Another dimension to the problem is the non-disclosure of hardware implementation details by processor vendors. This has two serious implications. Firstly, compilers are slow in adapting to architectural performance innovations. Even worse, in some cases legacy optimization techniques which performed well on previous hardware generations can actually perform poorly on newer hardware (for example, see the potential interactions of \emph{if-conversion} with branch prediction reported in \Cref{subsec:patterns_of_opts}). Secondly, programmers often have no clear view of the architecture's and compiler's internals and thus may produce code that is neither compiler- nor architecture-friendly. 

The goal of this paper is to provide systematic means of exposing optimization opportunities as part of the compiler's daily development cycle. Any methodology trying to achieve this should exhibit the following properties:

\begin{enumerate}
\item \emph{Portability}: The technique should be easily and rapidly portable across new versions of the same compiler and supported on all its target architectures.
\item \emph{Agility}: The overhead introduced by the discovery stage of the approach should be acceptable and contained within the day-to-day development cycle of compilers; for example, as part of the nightly testing system of a compiler.
\item \emph{Versatility}: To maximize productivity and outcome, it should be able to expose optimization patterns that can benefit a large number of programs and/or architectures rather than isolated cases.
\item \emph{Insightfulness}: The approach should correlate the observed behaviors of interest to their potential causes in a way that assists the compiler engineer in easily locating, understanding and acting upon the detected optimization opportunity.
\end{enumerate}

Iterative compilation and MLB techniques primarily focus on a given combination of application and compiler version to discover optimization sequences that can outperform the standard optimization levels of a compiler~\cite{Boyle:2018}. In their majority, they were not engineered to support the improvement of the standard optimization levels during the daily development cycle of a compiler.

While random iterative compilation has proven very successful in finding optimizations that can outperform the standard optimization levels~\cite{bodin:inria-00475919,CK:2018} for a selected application and architecture, it only complies with the \emph{portability} property; no major engineering is needed for porting to any modified version of the same compiler. The technique fails to address the rest of the desired properties since it is typically costly to run (hundreds to thousands of iterations need to be tested), and the random nature of it disallows the systematic discovery of optimization patterns and does not offer much information to assist the compiler developer in spotting the source of the optimization opportunities. On the other hand, while MLB techniques are able to expose optimization patterns and to some extent expose the underlying reasons for an optimization pattern, they usually fail to address the first two desired properties. More specifically, they are typically extremely costly to port to a new version of the same compiler, since any modification that affects the compiler's optimizer requires their expensive training phase to be repeated~\cite{Dubach:2009}. Since this training phase would need to be part of the daily compiler development cycle, they fail to exhibit the \emph{agility} property due to the associated overheads.

A nightly testing system repeatedly checks the ability of a compiler to optimize programs by tracking the performance of a set of benchmarks, in terms of execution time, code size, or even energy consumption, during the compiler's development phase. Mainstream compilers come with their own test suits, such as the LLVM Test Suite~\cite{llvm:test-suite}, which include a collection of programs that can be used as part of a nightly testing system. Nightly testing systems typically report performance degradations for each tested benchmark and between successive nightly testing sessions, but they provide no further means of identifying the causes of these degradations. Thus, the compiler engineer needs to go through a long discovery phase to find the changes that caused such degradations. More importantly, the current nightly routine testing provides no means of exposing new/hidden potential optimization opportunities, and of ranking these opportunities in terms of their potential impact. Since nightly testing is an essential part of the compiler's daily development cycle, it is the ideal place to provide means of optimizing the standard optimization levels of a compiler. Thus, this paper makes the following significant contributions:

\begin{itemize}
\item It proposes an enhanced nightly testing system that can help to tune the standard compiler optimization levels of a target-independent common optimizer to achieve better resource usage (execution time, energy consumption and code size);
\item It demonstrates how the newly introduced system can systematically expose hidden architecture-dependent and cross-architecture optimization opportunities (\Cref{sec:exposeOpts}).
\end{itemize}

In~\cite{Georgiou:2018:SCOPES}, we demonstrated that by performing fewer of the optimizations available in a standard compiler optimization level such as -O2 while preserving their original ordering, significant savings can be achieved in both execution time and energy consumption. This observation has been validated on embedded processors, namely the Arm Cortex-M0 and the Arm Cortex-M3, using two different versions of the LLVM compilation framework; v3.8 and v5.0. In this paper, we leverage the technique introduced in~\cite{Georgiou:2018:SCOPES} to devise a methodology which can enhance the nightly routine testing to enable the systematic tuning of the standard optimization levels and which exhibits all four desired properties defined earlier in this section: \emph{portability}, \emph{agility}, \emph{versatility} and \emph{insightfulness}.

To further evidence that the technique is easily portable to new versions of the same compiler and to new architectures, and thus, that it accommodates the \emph{portability} property, the technique was ported to a new version of the LLVM compiler, namely the LLVM 6.0.1, and applied to a broader class of architectures beyond the deeply embedded processors initially tested, i.e.\ the Intel i5-6300U X86-based architecture popular in desktop and laptop PCs, and the Arm Cortex-A53, an Armv8-A 64-bit based architecture frequently used in mobile devices (see~\Cref{sec:comp_and_analysis}). Both the compiler- and architecture-related porting were completed within an hour of engineering. 

While it is well known that optimization configurations better than the ones offered by standard optimization levels do exist in the complete optimization space~\cite{Boyle:2018}, we are interested in tuning the standard optimization levels. Thus, it is important to examine if tuning opportunities can consistently be found within the optimization space that the technique from~\cite{Georgiou:2018:SCOPES} exercises. This space consists of prefix subsequences of the optimization sequences applied by standard optimization levels. A formal definition of these optimization sequences is given in~\Cref{sec:comp_and_analysis}. Although we made an implicit choice of using only user-visible transformation passes of the optimizer for exploiting the optimization space, our approach is also capable of exposing the impact of pass parametrization, as demonstrated in~\Cref{subsec:patterns_of_opts}.

Experimental evaluation with 42 benchmarks belonging to the LLVM test suite~\cite{llvm:test-suite} demonstrated performance gains for at least half of the benchmarks, with an average of 11.5\% and 5.1\% execution time improvement over the standard optimization level -O3 for the i5-6300U and the Cortex-A53 processors, respectively. 
These findings confirm that the technique can detect optimization opportunities beyond deeply embedded architectures, like the Cortex-M0 and Cortex-M3 examined in~\cite{Georgiou:2018:SCOPES}, and across multiple versions of the LLVM compiler: namely the LLVM 3.8 and LLVM 5.0 used in~\cite{Georgiou:2018:SCOPES} and the LLVM 6.0.1 used in this paper.

The execution time for all the exploited configurations was in the range of several hours for each architecture. This is because only 64 and 66 configurations need to be tested for each benchmark for the Intel i5-6300U X86 and the Arm Cortex-A53 processors, respectively. As the introduced overhead is acceptable for a nightly regression system, our approach also exhibits the \emph{agility} property.

Our enhanced nightly testing then classifies the benchmarking results through a post-analysis that exposes architecture-dependent and cross-architecture optimization patterns, which fulfils the \emph{versatility} property.
This enhanced system, demonstrated in~\Cref{sec:exposeOpts}, directly pinpoints behaviors common across multiple benchmarks and architectures and reveals their possible causes, and thus, fulfils the \emph{insightfulness} property. Using only a selection of the information collected during the tests, we demonstrate the value of the enhanced system by exposing two significant cross-target shortcomings of the LLVM common optimizer, two distinct opportunities for target-aware heuristics adjustments and a possible direction for improving the support of advanced hardware branch prediction at compiler level. As a proof of concept, in~\Cref{subsec:caseStudy}, we implemented and tested one of the discovered tuning opportunities using the guidance provided by the enhanced nightly regression system. 
More specifically, by adding basic loop awareness to the CFG simplification pass of the LLVM optimizer, we were able to unlock the potential execution time reductions over the standard optimization level -O3 for the majority of the relevant test cases, and without affecting the performance of the rest of the benchmarks used.

The rest of the paper is organized as follows. \Cref{sec:comp_and_analysis} gives a brief overview of the common optimizer exploitation technique that was introduced in~\cite{Georgiou:2018:SCOPES} and the adjustments needed for the architectures used in this paper. Our benchmarking experimental evaluation results are presented and discussed in~\cref{subsec:benchmarking}. \Cref{sec:exposeOpts} introduces the enhanced nightly testing system and demonstrates how it can guide compiler engineers to tune the common compiler optimizer. \Cref{sec:related_work} critically reviews previous work related to ours. Finally, \Cref{sec:conc_future} concludes the paper and outlines opportunities for future work.

%%%%%%%%%%%%%%%%%%%%%%%%%%%%%%%%%%%% Compilation and Analysis %%%%%%%%%%%%%%%%%%%%%%%%%%%%%%%%%%%%%%%%%%%%%%

\section{Exploiting Standard Optimization Levels}
\label{sec:comp_and_analysis}

\begin{figure}[!t]
\centering
\includegraphics[trim={0.1cm 15.2cm 18cm 0cm}, scale=0.9 ,clip]{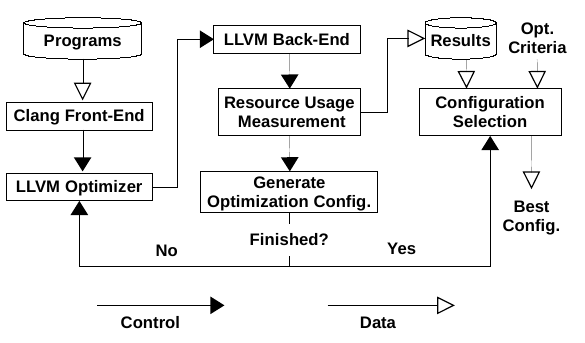}
\caption{Compilation and evaluation process (modified from \cite{Georgiou:2018:SCOPES}).}
\label{fig_evaluation_process}
\end{figure}

\Cref{fig_evaluation_process} demonstrates the process used to evaluate the effectiveness of the different optimization configurations to be explored. Each configuration is a sequence of flags used by the LLVM optimizer to drive the analysis and transformation passes.
An analysis pass can identify properties and expose optimization opportunities that can later be used by transformation passes to perform optimizations. 
A standard optimization level (\mbox{-O1}, \mbox{-O2}, \mbox{-O3}, \mbox{-Os}, \mbox{-Oz}) can be selected as the starting point. Each optimization level represents a list of transformation and analysis flags. 
The order of these flags defines the order in which the transformation and analysis passes will be applied to the code under compilation. A new flag configuration is obtained from the current list of flags by removing the last transformation flag together with all preceding analysis flags up to but not including the next found transformation flag.
Then the new optimization configuration is being applied to the unoptimized IR of the program, obtained from the Clang front-end. In this way, the program's unoptimized IR only needs to be generated once by the Clang front-end; it can then be used throughout the exploration process, thus saving compilation time. 
The optimized IR is then passed to the LLVM back-end and linker to generate the executable for the architecture under consideration. Note that both the back-end and linker are always called using the optimization level selected for exploration, in our case \mbox{-O3}. The executable's resource usage is measured and stored for each tested configuration. The exploration process finishes when the current list of transformation flags is empty. 
This corresponds to the optimizer applying no optimizations, i.e.\ the IR is left unoptimized. A more detailed explanation of the technique is given in~\cite{Georgiou:2018:SCOPES}.

More formally, the generation of the optimization configurations can be defined as follows:
\smallskip

Let $S$ be the sequence of optimization flags obtained from the selected standard optimization level
\begin{equation}
    S = \langle F_1,F_2,...,F_n\rangle.
\label{eq:S}
\end{equation}

\noindent
$S$ includes flags $F_i$, $1\leq i \leq n$, with $n$ being the total number of flags appearing in the selected optimization level. Let $F^T$ denote the set of transformation passes and $F^A$ the set of analysis passes, with \mbox{$F^T \cap F^A = \emptyset$}. Thus, each flag in $S$ is either a transformation pass, \mbox{$F_i \in F^T$}, or an analysis pass, \mbox{$F_i \in F^A$}. The set of generated optimization configurations, $G$, is defined as:
\begin{equation}
    G = \{ \hat{S}_0, \hat{S}_1, ..., \hat{S}_t \} 
\label{eq:G}
\end{equation}
where $t$ is the total number of transformation flags appearing in $S$.
$\hat{S}_0$ represents the configuration where no transformation or analysis passes are being applied by the common optimizer, 
and $\{ \hat{S}_1, ..., \hat{S}_t \}$ is the set of all prefixes $\hat{S}_i$ of $S$, $0<i\leq t$, that end in an optimization flag, i.e.\ 
\begin{equation}
%\begin{split}
\{\hat{S}_i \sqsubseteq S \ |\  \hat{S}_i = \langle F_1,...,F_k\rangle \ \wedge\ F_k \in F^T \}.
%\end{split}
\label{eq:Si}
\end{equation}

This method of generating the optimization configurations ensures {\em by construction} that all configurations are prefixes of the standard optimization level targeted for tuning. This has the following three advantages:

\begin{enumerate}
\item The new configurations do not trigger the LLVM optimizer's pass manager to enable additional transformation passes. 
This is because within the generated prefix configurations of a standard optimization level, the knowledge built into the optimization level regarding effective pass orderings is preserved. 
We have manually verified this by using the \emph{-debug-pass=Executions} flag while running the LLVM optimizer, which emits the sequence of passes executed by the common optimizer.

\item Using optimization configurations that are prefixes of the standard optimization levels should result in the generation of valid executables, because the standard optimization levels of each new version of a compiler are typically heavily tested to ensure the production of functionally correct executables. If not, this might be an indication of a bug in the optimizer.

\item The systematic generation of optimization configurations, as opposed to using random selection, results in a meaningful relation of the generated configurations to the exploited standard optimization level. This is critical for exposing optimization patterns and enables the discovery and understanding of the potential optimization causes, as demonstrated in \Cref{sec:exposeOpts}. Thus, both, \emph{versatility} and \emph{insightfulness} from \Cref{sec:introduction} can be achieved.
\end{enumerate}

In~\cite{Georgiou:2018:SCOPES}, the primary focus was deeply-embedded processors, typically used in Internet of Things (IoT) applications, and thus, we demonstrated the technique's effectiveness on the Arm Cortex-M0 \cite{Cortex_M0} and the Arm Cortex-M3 \cite{Cortex_M3} processors. In this paper, the technique is being ported to two more complex processors, namely the Intel i5-6300U and the Arm Cortex-A53. Porting to a new architecture is not time-consuming since the technique treats an architecture as a black box. This is feasible because no resource models are required, neither for execution time nor for energy consumption. Instead, physical measurements of such resources can be used to assess the effectiveness of a new optimization configuration on a program. 

Similarly, the technique treats the compiler as a black box. It only uses the compilation framework to exercise the different optimization configuration scenarios extracted from a predefined optimization level on a particular program. In contrast, machine-learning-based techniques typically require a heavy training phase for each new compiler version or when a new optimization flag is introduced~\cite{Ashouri:2017, blackmore2015}. Considering the range of the LLVM versions the technique was applied to, a total of two years of compiler developments between LLVM versions v3.8, used in~\cite{Georgiou:2018:SCOPES}, and v6.0.1, used in this paper, we can safely assume the \emph{portability} of the technique across different versions of the same compiler. In fact, we can argue that our technique can be ported to any compiler optimizer, provided that its optimization process depends on sequences of optimization passes, and that it exposes means of controlling these subsequences. Overall, the porting to the new architectures and the new version of the LLVM compiler was completed within an hour. 

The Collective Knowledge (CK)~\cite{CK:2018,CK:online}, a framework for collaborative research that supports compilers' optimization auto-tuning, was used for evaluation. CK includes a variety of benchmark suites for the training and the evaluation of auto-tuning techniques for compiler optimization, such as iterative-based and ML-based techniques. One of them is the Milepost-GCC-Codelet benchmark suite, which was used in the seminal work on ML-based compiler optimization, MilePostGCC~\cite{Fursin2011}. These benchmarks represent hot spots extracted together with there datasets from several real software projects~\cite{CK:2018}. These benchmarks are also part of the LLVM compiler test-suite, under the MiBench benchmark suite~\cite{llvm:test-suite}. Both the Milepost-GCC and its benchmark suite are now integrated into the CK framework and are often used as the baseline to compare the effectiveness of new auto-tuning techniques~\cite{CK:2018,Ashouri:2017,blackmore2015}. Thus, this paper also uses the Milepost-GCC-Codelet for evaluation. 
Although these benchmarks can be considered small, with between 20 and 270 lines of source code, using small benchmarks is evidently sufficient to spot potential optimization opportunities, as demonstrated later on in~\Cref{sec:exposeOpts}. This has two major advantages: it simplifies the understanding of the performance variations across different executables of the same benchmark, and, in addition, the compilation, measurement and validation cycles are kept short.

The Resource Usage Measurement box that is part of our compilation and evaluation process, shown in~\cref{fig_evaluation_process}, can be used to determine the execution time, energy consumption and code size for each executable generated. Various measurement or estimation techniques can be utilized as part of our framework in a plug-and-play approach to address different hardware platforms and optimization requirements. For this work, we focus on the execution time and code size since there is no on-chip energy measurement support for the CortexA-53 processor under test. As demonstrated in~\cite{Georgiou:2018:SCOPES}, the technique is capable of accounting for energy consumption, when sufficiently accurate energy measurements or estimations are available. The code size can be obtained by examining the size of the ``.text'' section of the executable.

For execution time measurements we use the CK's built-in execution time measurement framework. The framework has a calibration process that is needed prior to measurement to determine the number of times a benchmark should be executed in a loop while measuring to obtain a representative average execution time for each benchmark. 
In addition, we repeated the evaluation process ten times and excluded the two highest and two lowest execution time values for each benchmark. This minimizes the impact of other events on the benchmark's execution time, such as Dynamic Voltage and Frequency Scaling (DVFS) and noise from the operating system or other applications running on the machines under test. In the case of the Intel i5-6300U, the above procedure was adequate to provide stable and trustworthy results. In the case of the Arm Cortex-A53, as used in a Raspberry Pi 3B+ board, the measurements were still not sufficiently stable.

Further investigation showed that the overclocking functionality of the processor that allows it to go beyond its maximum specified factory frequency, from 1200MHz to 1400MHz, is counterproductive as the board has no means of proper cooling and was frequently overheating, reaching temperatures above those in its specifications. This was causing the processor to throttle and exhibit unstable behavior. Restricting the maximum frequency to 1200MHz, we repeated our measurements, allowing DVFS. The measurement variation was now within a [-4.5\%, 4.5\%] range. Although this variation is acceptable and the compiler optimization effects can be identified in most cases, we decided to follow the official LLVM guidelines for benchmarking compilers, which suggest disabling DVFS and modules external to the processor~\cite{LLVM_tips:online}. This helps to isolate the effect of the compiler optimizations on performance from other artefacts that can affect the execution time, such as DVFS. Hence, DVFS was disabled, fixing the processor's frequency to 1200 MHz, and the wireless communication (WiFi and Bluetooth) modules were also disabled.
Finally, the mean was obtained over the remaining six out of ten measured execution time values; all of the six values were within the [-1\%,1\%] range from the mean value. We also evaluated a faster approach in which only five repeated measurements were taken, the two extreme values being discarded, and the arithmetic mean of the remaining three measurements was used as the representative value.  The retained values were contained within the [-1.5\%, 1.5\%] range from the arithmetic mean for 99\% of all tested configurations, and the potential throughput of the test platform was doubled compared to the ten-fold repetition of measurements. 

CK has a built-in self-test mechanism that detects and reports when a generated executable is invalid, i.e.\, it does not provide the expected results. We modified this mechanism to check the benchmark's results for each optimization configuration against those of the \hbox{-O0} compilation with no optimizations enabled. This is because compiled unoptimized programs are considered reference models that act as intended by the programmer.

%%%%%%%%%%%%%%%%%%%%%%%%%%%%%%%%%%%%%%%%%%%%%%%%%%%% RESULTS %%%%%%%%%%%%%%%%%%%%%%%%%%%%%%%%%%%%%%%%%%%%%%%%%%%%%%%%%%%%%
\section{Benchmark Evaluation}
\label{subsec:benchmarking}

\begin{figure*}[!htp]
    \centering
    \begin{subfigure}{\textwidth}
    \centering
      \includegraphics[trim=0.25cm 0.25cm 0.25cm 0.3cm,clip=true,width=0.8\linewidth]{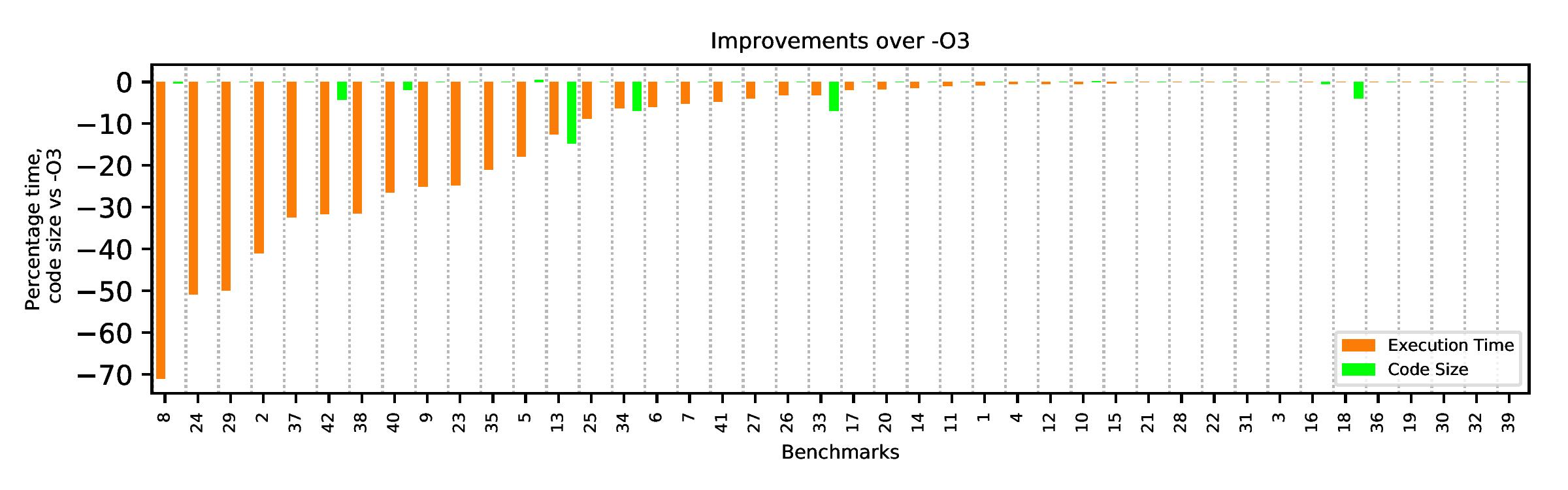}
      \subcaption{Results for the i56300U processor and the LLVM v6.0.1 compilation framework.}
      \label{subfig:i5-6300U_total_result}
    \end{subfigure} %\hfill
    \begin{subfigure}{\textwidth}
    \centering
        \includegraphics[trim=0.25cm 0.25cm 0.25cm 0.3cm,clip=true,width=0.8\linewidth]{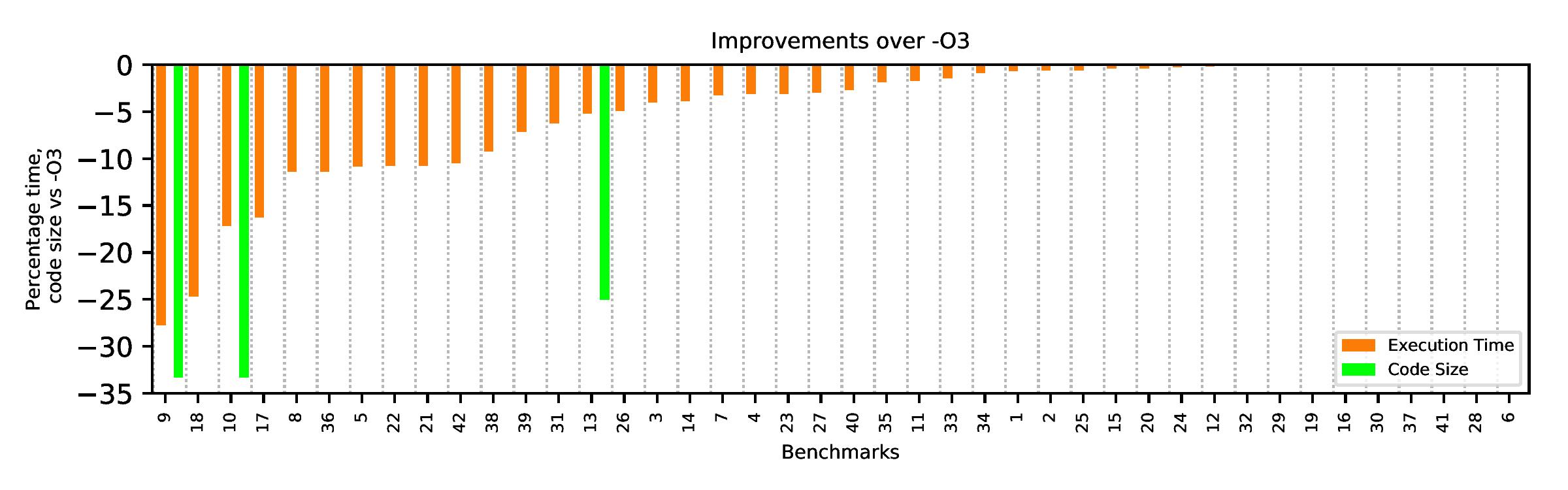}
      \subcaption{Results for the Cortex-A53 processor and the LLVM v6.0.1 compilation framework.}
      \label{subfig:CortexA53_total_result}
    \end{subfigure}

  \begin{subtable}{\textwidth}
    \vspace{0.3cm}
  \centering
  \footnotesize

  \begin{tabular}{|>{}c|>{}l|>{}l|>{}c|>{}l|}
  \hline

    \textbf{ID} & \multicolumn{2}{c|}{\textbf{Benchmark Name}} & \textbf{ID} & \multicolumn{1}{c|}{\textbf{Benchmark Name}} \\ \hline
    1 & \multicolumn{2}{>{}l|}{automotive-basicmath-cubic-3-1} &      2 & automotive-basicmath-isqrt-1-1 \\ \hline
    3 & \multicolumn{2}{>{}l|}{automotive-qsort1-src-qsort-1-1} &       4 & automotive-susan-e-src-susan-10-1 \\ \hline
    5 & \multicolumn{2}{>{}l|}{automotive-susan-e-src-susan-2-1} &      6 & automotive-susan-s-src-susan-1-1 \\ \hline
    7 & \multicolumn{2}{>{}l|}{consumer-jpeg-c-src-jcdctmgr-13-1} &     8  & consumer-jpeg-c-src-jchuff-9-1 \\ \hline
    9  & \multicolumn{2}{>{}l|}{consumer-jpeg-c-src-jfdctint-2-1} &     10 & consumer-lame-src-fft-2-1 \\ \hline
    11 & \multicolumn{2}{>{}l|}{consumer-lame-src-newmdct-10-1} &       12 & consumer-lame-src-newmdct-3-1 \\ \hline
    13 & \multicolumn{2}{>{}l|}{consumer-lame-src-psymodel-17-1} &      14 & consumer-lame-src-quantize-7-1 \\ \hline
    15 & \multicolumn{2}{>{}l|}{consumer-lame-src-quantize-pvt-6-1} &     16 & consumer-lame-src-takehiro-16-1 \\ \hline
    17 & \multicolumn{2}{>{}l|}{consumer-lame-src-takehiro-5-1} &       18 & consumer-mad-src-layer3-5-1 \\ \hline
    19 & \multicolumn{2}{>{}l|}{consumer-mad-src-layer3-6-1} &        20 & consumer-tiff2rgba-src-tif-predict-4-1 \\ \hline
    21 & \multicolumn{2}{>{}l|}{consumer-tiffdither-src-tif-fax3-8-1} &   22 & consumer-tiffdither-src-tif-fax3-9-1 \\ \hline
    23 & \multicolumn{2}{>{}l|}{consumer-tiffdither-src-tiffdither-1-1} &   24 & consumer-tiffmedian-src-tiffmedian-1-1 \\ \hline
    25 & \multicolumn{2}{>{}l|}{consumer-tiffmedian-src-tiffmedian-3-1} &   26 & consumer-tiffmedian-src-tiffmedian-4-1 \\ \hline
    27 & \multicolumn{2}{>{}l|}{consumer-tiffmedian-src-tiffmedian-5-1} &   28 & consumer-tiffmedian-src-tiffmedian-6-1 \\ \hline
    29 & \multicolumn{2}{>{}l|}{network-dijkstra-src-dijkstra-large-5-1} &  30 & office-ghostscript-src-gdevpbm-1-1 \\ \hline
    31 & \multicolumn{2}{>{}l|}{office-rsynth-src-nsynth-5-1} &       32 & office-rsynth-src-nsynth-9-1 \\ \hline
    33 & \multicolumn{2}{>{}l|}{security-pgp-d-src-mpilib-1-1} &      34 & security-pgp-e-src-mpilib-1-1 \\ \hline
    35 & \multicolumn{2}{>{}l|}{security-pgp-e-src-mpilib-3-1} &      36 & security-pgp-e-src-mpilib-4-1 \\ \hline
    37 & \multicolumn{2}{>{}l|}{telecomm-adpcm-c-src-adpcm-1-1} &       38 & telecomm-adpcm-d-src-adpcm-1-1 \\ \hline
    39 & \multicolumn{2}{>{}l|}{telecomm-fft-fftmisc-5-1} &         40 & telecomm-fft-fourierf-3-1 \\ \hline
    41 & \multicolumn{2}{>{}l|}{telecomm-gsm-src-rpe-4-1} &         42 & telecomm-gsm-src-short-term-2-1 \\ \hline
  \end{tabular}%
  \subcaption{The GCC-Milepost benchmarks used for evaluation.}
    \label{subtab:benchs_table}
  \end{subtable}
    \caption{Best achieved execution time improvements over the standard optimization level -O3. For the best execution time optimization configuration, code size improvements are also given. A negative percentage represents a reduction of resource usage compared to -O3.}
    \label{fig:all_results_graphs}
\end{figure*}

The 42 benchmarks from the CK Milepost-GCC-Codelet benchmark suite, listed in Table \ref{subtab:benchs_table}, were used for both the Intel i5-6300U and the Arm Cortex-A53 processors to facilitate the discovery of potential cross-architecture compiler optimizations. For each benchmark, \Cref{fig:all_results_graphs} (\Cref{subfig:i5-6300U_total_result} for the i5-6300U and \Cref{subfig:CortexA53_total_result} for the Cortex-A53) demonstrates the biggest performance gains achieved by the proposed technique compared to the standard optimization level under investigation, -O3. In other words, this figure represents the resource usage results of the optimization configuration which achieved the best performance gains among the configurations exercised by our technique, when compared to -O3 for each benchmark. A negative percentage represents an improvement on a resource, e.g., a result of -20\% for execution time represents a 20\% reduction in the execution time obtained by the selected optimization configuration when compared to the execution time of the reference -O3 optimization configuration.

The code size improvements are also given for the selected configurations to demonstrate that our technique can affect code size. However, code size is only relevant for embedded architectures, and, thus, is not investigated further in this paper. When dealing with deeply embedded architectures, such as the Cortex-M0 and Cortex-M3 examined in~\cite{Georgiou:2018:SCOPES}, code size is often the first resource targeted for optimization due to the limited memory of the processor. In such cases, our optimization exploitation can use the -Os or -Oz optimization levels as a starting point; both aim to achieve smaller code size.
The optimization criteria, and thus the optimization level used as a starting point, can be altered according to the resource requirements for a specific application. Energy consumption can be another resource to be exploited whenever accurate energy measurements are available for the processor under investigation.

For the i5-6300U processor, we observed an average reduction in execution time of 11.5\%,~with 26 out of the 42 benchmarks seeing execution time improvements over -O3 ranging from around 1\% to 71\%.~For the Cortex-A53 processor, we observed an average reduction in execution time of 5.1\%, with 26 out of the 42 benchmarks seeing execution time improvements over -O3 ranging from around 1\% to 28\%.

\begin{figure*}[!htp]
 \centering
\includegraphics[trim=0.25cm 0.25cm 0.25cm 0.1cm,clip=true,width=\linewidth]{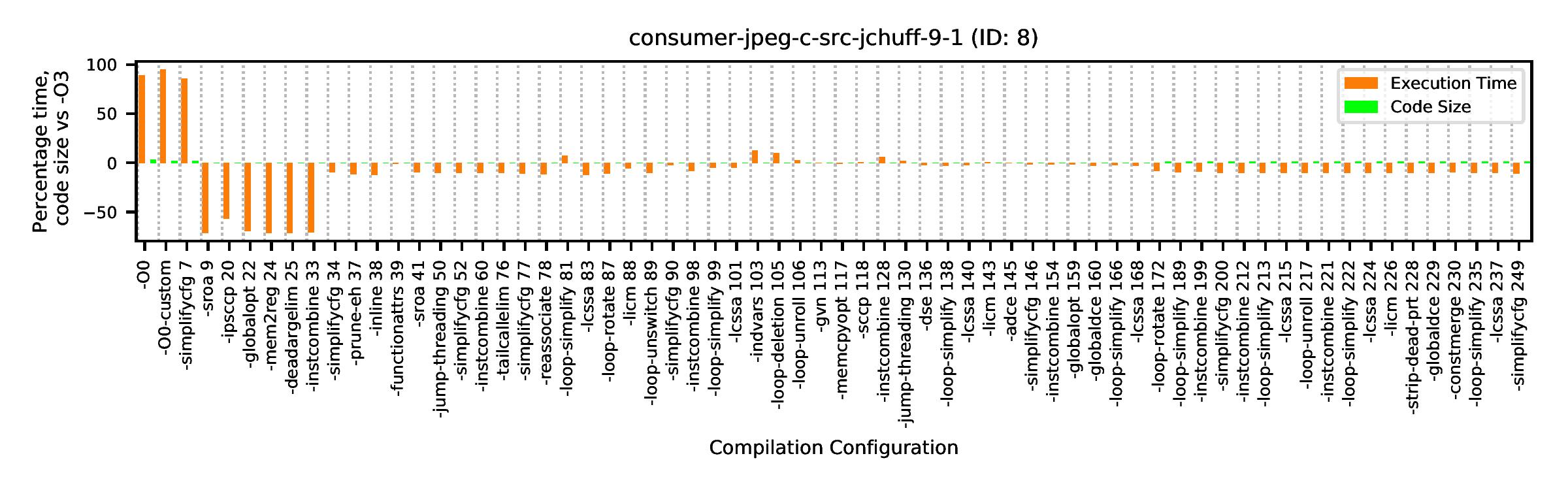}
\caption{Optimization-performance example on the i5-6300U. For each optimization configuration tested by the proposed technique, the execution time and code size improvements over -O3 are given. A negative percentage represents a reduction of resource usage compared to -O3. Each element of the horizontal axis has the name of the last flag applied and the total number of flags used. A description of the optimization flags used can be found at~\cite{LLVM:passes2020}. The configurations are prefixes of the -O3 optimization sequence, starting from -O0 and adding optimization flags in order of occurrence till reaching the complete -O3 sequence of flags.}
\label{fig:demo_figure}
\end{figure*}

\Cref{fig:demo_figure} demonstrates the effect of each optimization configuration, exercised by our exploitation technique, on the two resources (execution time and code size), for the \emph{consumer-jpeg-c-src-jchuff-9-1} benchmark on the i5-6300U processor. Similar figures were obtained for all the 42 benchmarks and for both of the processors. As in \Cref{fig:all_results_graphs}, a negative percentage represents a reduction (thus, an improvement) in the usage of the given resource compared to the one achieved by standard \mbox{-O3} optimization. The horizontal axis of the figures shows the flag at which compilation stopped together with the total number of flags included up to that point. This represents an optimization configuration that is a prefix of the -O3 optimization sequence. For example, the best optimization configuration for performance for the benchmark in \Cref{fig:demo_figure} is achieved when the compilation stops at flag number 9, \emph{sroa} (static replacement of
aggregates). This means that the optimization configuration includes the first nine flags of the -O3 configuration with their original ordering preserved. The optimization configurations include both transformations and analyses passes. The \mbox{\emph{-O0-custom}} configuration is the split version of the -O0 optimization level where the compilation is explicitly decomposed into the front-end, common optimizer and back-end, as described in \Cref{sec:comp_and_analysis}. Its results are not equivalent to the normal -O0 compilation, since the back-end is driven by the -O3 optimization level.

The number of optimization configurations exercised in each case depends on the number of transformation flags included in the -O3 level of the version of the LLVM optimizer used. Note that we are only considering the transformation passes visible to the compiler user~\cite{LLVM:passes2020} and do not explore the features that are explicitly hidden from the user, such as various transformation parameters. Using this approach, 64 and 66 different configurations are being automatically detected and tested by our technique for the Cortex-A53 and the i5-6300U processors, respectively. The difference for the \mbox{-O3} optimization level in terms of optimization flags between the two processors is probably an attempt by the compiler engineers to better address the performance characteristics of the two architectures. Overall, more analysis passes are used for the i5-6300U processor. Many of the transformation passes are applied multiple times in a standard optimization level, but because of their different position in the configuration sequence they may have a different effect. Thus, we consider each repetition as an opportunity to create a new optimization configuration. 
Furthermore, note that more, hidden to the user, transformation passes exist in the LLVM optimizer, but typically, these are passes that have implicit dependencies on the passes that are exposed via compilation flags. The methodology of creating a new optimization configuration, explained in \Cref{sec:comp_and_analysis}, does not affect such implicit dependencies. This is because the LLVM pass manager ensures that such dependencies are always met.

It is time-consuming to identify any optimization patterns across multiple benchmarks, by manually inspecting the compilation profiles obtained for all the benchmarks, similar to the ones presented in \Cref{fig:demo_figure}. 
In the next section, we show how the benchmarks can be automatically clustered based on their compilation profiles. We then demonstrate the value of such clustering as part of a nightly testing system, as it can expose potential hidden architecture-dependent and cross-architecture optimizations. Moreover, it can pinpoint optimizations that degrade performance.

%%%%%%%%%%%%%%%%%%%%%%%%%%%%%%%%%%%%%%%%%%%%%%%%%% Exposin Hidden OPTS %%%%%%%%%%%%%%%%%%%%%%%%%%%%%%%%%%%%%%%%%%%%%%%

\section{Exposing Hidden Optimization Opportunities}
\label{sec:exposeOpts}

Retargetable compiler frameworks achieve their generality by abstracting target architecture properties and by relying on cross-target heuristics in the front- and middle-end compilation passes. The abstract properties may be parametrized by quantitative characteristics of each actual target used, but the decision heuristics and the actual sequence of optimizations are often defined by experimentation and, once established, are seldom questioned in subsequent releases of the compiler framework. Therefore, evaluating the pertinence and the quality of the heuristics used in a compiler may provide valuable insights into the quality of the current compiler configuration and its potential for further improvement.

The standard approach to tuning a compiler's common optimizer remains the repetitive testing of the compiler on a variety of benchmarks and mainstream architectures. This approach is typically called nightly routine testing, and it mainly aims at validating benchmark results in terms of correctness and improving performance (or, in some application domains, code size). The output of a nightly testing session is typically a report with information about the compilation time, execution time and the correctness of the output for each test. These results are then compared to a reference point, usually the result of a previous nightly testing run that passed all the validation tests and exhibits the best achievable execution and compilation times so far. The purpose of nightly testing is to constantly monitor the quality of the modifications in a compiler towards the release of a new version.

All observed regressions (either correctness failures or significant degradations in the execution time of a benchmark relative to its reference point) have to be investigated by a compiler engineer. However, the detection of a regression does not offer any insights into what actually caused it and requires the engineer to manually examine and track the source of the problem. Depending on the engineer's experience and the complexity of the issue, the identification of the root cause of a regression can be an extremely time-consuming task. Furthermore, a standard nightly testing system will only report regressions or improvements for individual tests, but will not directly pinpoint behaviors common across multiple benchmarks and architectures that can indicate hidden optimization opportunities.

In this section we propose an enhancement of the classic nightly testing that utilizes the technique explained in \Cref{sec:comp_and_analysis} to extract {\em recurring} behaviors of the compiler. By exposing and quantifying the effect of successive optimizations across all tested benchmarks and supported target architectures of a compiler, the enhanced nightly testing approach enables the discovery of unexploited cross-architecture and architecture-dependent optimization opportunities and the identification of optimization passes that have a negative impact on target resource utilization (execution time, energy consumption, or code size). The insights gained in this way can drastically improve the process of tuning the compiler's common optimizer, even without detailed knowledge of the target architecture.

To demonstrate this, we will use the results obtained by our technique on the Milepost-GCC benchmarks, as described in \Cref{subsec:benchmarking}. Milepost-GCC benchmarks are an excellent candidate for this exercise as they are also part of the LLVM compiler's test suite (under the MiBench subsuite \cite{llvm:test-suite}). From \Cref{fig:all_results_graphs}, we already know that significant performance gains can be achieved using the proposed technique across both architectures. A compiler engineer will need to focus first on the cases where the same optimizations appear across multiple benchmarks. These repeating patterns indicate potential optimization opportunities that can benefit a wider group of programs and/or architectures. To this end, the benchmark results are first classified to expose common optimization behaviors which are then analyzed in more depth.

\subsection{Classification of nightly testing results}
\label{subsec:resultClassification}
\Cref{fig:optsFrequencies} shows the outcome of the initial result classification. \Cref{subfig:i5-regression-report} and \Cref{subfig:rpi3Bplus-regression-report} are the new proposed reports for a compiler's nightly testing system for the i5-6300U and the Cortex-A53 processors, respectively. The reports include all the benchmarks where our technique achieved an execution time reduction of more than 3\%. This threshold can be adjusted by compiler engineers to allow them to focus on the cases for which they consider that the execution time reduction is sufficient to warrant further investigation. The benchmarks are then grouped in terms of their observed optimization behavior. The first level of grouping is done on the \emph{First Config.\ Better than -O3} column, which represents the first optimization configuration that outperformed the -O3 (e.g., pass \emph{sroa 9} in \Cref{subfig:i5:simplifycfg34}), and on the \emph{Config.\ Removing Gains} column, which represents the configuration in which those achieved gains were lost by the addition of more optimization flags (e.g., pass \emph{simplifycfg 34} in \Cref{subfig:i5:simplifycfg34}). The second grouping appears on the \emph{Best Overall Config.}\ column which represents the configuration that achieved the best performance against -O3. Finally, the benchmarks within groups are sorted based on their achieved performance gains over -O3, in descending order. This is also the case for any benchmarks that do not belong to any group, e.g., the last 8 benchmarks in \Cref{subfig:rpi3Bplus-regression-report}. The reports presented in \Cref{fig:optsFrequencies} will be used in the later sections to demonstrate how they can guide the tuning of the compiler's optimizer.

\begin{figure*}[!htp]
    \centering
    \begin{subfigure}{\textwidth}
    \centering
    \footnotesize
  \resizebox{\textwidth}{!}{%
  \begin{tabular}{|c|c|c|c|r|}
  \hline
  \textbf{Benchmark ID} & \textbf{First Config.\ Better than -O3}   & \textbf{Config.\ Removing Gains}          & \textbf{Best Overall Config.}            & \textbf{Execution Time Reduction \%} \\ \hline
  8                     & \cellcolor[HTML]{96FFFB}sroa  - 9        & \cellcolor[HTML]{96FFFB}simplifycfg - 34 & \cellcolor[HTML]{67FD9A}instcombine - 33 & \cellcolor[HTML]{FFFFFF}-70.98       \\ \hline
  2                     & \cellcolor[HTML]{96FFFB}sroa  - 9        & \cellcolor[HTML]{96FFFB}simplifycfg - 34 & \cellcolor[HTML]{67FD9A}instcombine - 33 & \cellcolor[HTML]{FFFFFF}-40.98       \\ \hline
  37                    & \cellcolor[HTML]{96FFFB}sroa  - 9        & \cellcolor[HTML]{96FFFB}simplifycfg - 34 & \cellcolor[HTML]{67FD9A}instcombine - 33 & \cellcolor[HTML]{FFFFFF}-32.34       \\ \hline
  23                    & \cellcolor[HTML]{96FFFB}sroa  - 9        & \cellcolor[HTML]{96FFFB}simplifycfg - 34 & \cellcolor[HTML]{67FD9A}instcombine - 33 & \cellcolor[HTML]{FFFFFF}-24.76       \\ \hline
  13                    & \cellcolor[HTML]{96FFFB}sroa  - 9        & \cellcolor[HTML]{96FFFB}simplifycfg - 34 & \cellcolor[HTML]{F56B00}sroa  - 9        & \cellcolor[HTML]{FFFFFF}-12.53       \\ \hline
  25                    & \cellcolor[HTML]{96FFFB}sroa  - 9        & \cellcolor[HTML]{96FFFB}simplifycfg - 34 & \cellcolor[HTML]{F56B00}sroa  - 9        & \cellcolor[HTML]{FFFFFF}-8.82        \\ \hline
  7                     & \cellcolor[HTML]{96FFFB}sroa  - 9        & \cellcolor[HTML]{96FFFB}simplifycfg - 34 & \cellcolor[HTML]{F56B00}sroa  - 9        & \cellcolor[HTML]{FFFFFF}-5.11        \\ \hline
  42                    & \cellcolor[HTML]{96FFFB}sroa  - 9        & \cellcolor[HTML]{96FFFB}simplifycfg - 34 & ipsccp -20                               & -31.61                               \\ \hline
  35                    & \cellcolor[HTML]{96FFFB}sroa  - 9        & \cellcolor[HTML]{96FFFB}simplifycfg - 34 & instcombine - 221                        & -21.05                               \\ \hline
  24                    & \cellcolor[HTML]{96FFFB}sroa  - 9        & \cellcolor[HTML]{96FFFB}simplifycfg - 90 & functionattrs - 39                       & -50.79                               \\ \hline
  34                    & \cellcolor[HTML]{F8FF00}instcombine - 33 & \cellcolor[HTML]{F8FF00}lcssa - 83       & \cellcolor[HTML]{F8FF00}instcombine - 33 & \cellcolor[HTML]{FFFFFF}-6.25        \\ \hline
  33                    & \cellcolor[HTML]{F8FF00}instcombine - 33 & \cellcolor[HTML]{F8FF00}lcssa - 83       & \cellcolor[HTML]{F8FF00}instcombine - 33 & \cellcolor[HTML]{FFFFFF}-3.13        \\ \hline
  29                    & no pattern                               & no pattern                               & jump-threading - 130                     & -50.00                               \\ \hline
  38                    & sroa  - 9                                & instcombine - 60                         & instcombine - 33                         & -31.53                               \\ \hline
  40                    & sroa  - 9                                & loop-rotate - 87                         & ipsccp -20                               & -26.51                               \\ \hline
  9                     & loop-unroll - 217                        & after simplifycfg -249                   & mem2reg - 24                             & -25.00                               \\ \hline
  5                     & no pattern                               & no pattern                               & loop-simplify 138                        & -17.82                               \\ \hline
  6                     & sroa  - 9                                & globaldce - 229                          & loop-rotate - 87                         & -6.00                                \\ \hline
  41                    & reassociate - 78                         & indvars - 103                            & loop-rotate - 87                         & -4.76                                \\ \hline
  27                    & sroa  - 9                                & lcssa - 101                              & ipsccp -20                               & -3.92                                \\ \hline
  26                    & loop-rotate - 87                         & instcombine - 98                         & loop-rotate - 87                         & -3.17                                \\ \hline
  \end{tabular}%
  }

  \subcaption{Advanced nightly testing report for the i5-6300U processor.} 
    \label{subfig:i5-regression-report}
    \end{subfigure}\hfill
    \vspace{.3cm}
\begin{subfigure}{\textwidth}
\centering
\footnotesize

  \centering
  \resizebox{\textwidth}{!}{%
  \begin{tabular}{|c|c|c|c|r|}
  \hline
  \textbf{Benchmark ID} & \textbf{First Config.\ Better than -O3}     & \textbf{Config.\ Removing Gains}              & \textbf{Best Overall Config.} & \textbf{Execution Time Reduction \%} \\ \hline
  10                    & \cellcolor[HTML]{9698ED}sroa - 8          & \cellcolor[HTML]{9698ED}instcombine - 27     & \cellcolor[HTML]{9698ED}sroa - 8     & -17.18                               \\ \hline
  36                    & \cellcolor[HTML]{9698ED}sroa - 8          & \cellcolor[HTML]{9698ED}instcombine - 27     & \cellcolor[HTML]{9698ED}sroa - 8     & -11.35                               \\ \hline
  42                    & \cellcolor[HTML]{9698ED}sroa - 8          & \cellcolor[HTML]{9698ED}instcombine - 27     & \cellcolor[HTML]{9698ED}sroa - 8     & -10.48                               \\ \hline
  31                    & \cellcolor[HTML]{9698ED}sroa - 8          & \cellcolor[HTML]{9698ED}instcombine - 27     & \cellcolor[HTML]{9698ED}sroa - 8     & -6.25                                \\ \hline
  7                     & \cellcolor[HTML]{9698ED}sroa - 8          & \cellcolor[HTML]{9698ED}instcombine - 27     & \cellcolor[HTML]{9698ED}sroa - 8     & -3.23                                \\ \hline
  5                     & \cellcolor[HTML]{9AFF99}loop-rotate - 73   & \cellcolor[HTML]{9AFF99}jump-threading - 109 & \cellcolor[HTML]{FD6864}instcombine - 80    & -10.82                               \\ \hline
  22                    & \cellcolor[HTML]{9AFF99}loop-rotate - 73   & \cellcolor[HTML]{9AFF99}jump-threading - 109 & \cellcolor[HTML]{FD6864}instcombine - 80    & -10.71                               \\ \hline
  21                    & \cellcolor[HTML]{9AFF99}loop-rotate - 73   & \cellcolor[HTML]{9AFF99}jump-threading - 109 & memcopyopt - 100                            & -10.71                               \\ \hline
  39                    & \cellcolor[HTML]{999903}loop-rotate - 73   & \cellcolor[HTML]{999903}instcombine - 80     & loop-rotate - 73                            & -7.14                                \\ \hline
  26                    & \cellcolor[HTML]{999903}loop-rotate - 73   & \cellcolor[HTML]{999903}instcombine - 80     & simplifycfg - 76                            & -4.92                                \\ \hline
  13                    & \cellcolor[HTML]{329A9D}loop-unswitch - 75 & \cellcolor[HTML]{329A9D}instcombine - 80     & loop-unswitch - 75                            & -5.16                                \\ \hline
  23                    & \cellcolor[HTML]{329A9D}loop-unswitch - 75 & \cellcolor[HTML]{329A9D}instcombine - 80     & simplifycfg - 76                         & -3.07                                \\ \hline
  9                     & loop-rotate - 145                          & loop-unroll - 186                            & loop-simplify - 182                            & -27.72                               \\ \hline
  18                    & loop-rotate - 145                          & no pattern                                   & strip-dead-prot - 194                            & -24.68                               \\ \hline
  17                    & sroa - 8                                  & loop-rotate - 73                             & ipsccp - 19                             & -16.23                               \\ \hline
  8                     & sroa - 8                                  & instcombine - 80                             & globalopt - 20                             & -11.38                               \\ \hline
  38                    & sroa - 8                                  & instcombine - 53                             & sroa - 8                             & -9.22                                \\ \hline
  3                     & no pattern                                 & no pattern                                   & licm - 192                            & -4.00                                   \\ \hline
  14                    & sroa - 8                                  & indvars - 86                                 & functionattrs - 33                            & -3.85                                \\ \hline
  4                     & no pattern                                & no pattern                                   & strip-dead-prot - 194                            & -3.07                                \\ \hline
  \end{tabular}%
  }
  \subcaption{Advanced nightly testing report for the Cortex-A53 processor.}
  \label{subfig:rpi3Bplus-regression-report}

\end{subfigure} \hfill
\caption{Advanced regression reports using our technique on the Milepost-GCC benchmarks. 
  The colors represent benchmarks' grouping based on their optimization behavior. The first level of grouping is done on the \emph{First Config.\ Better than -O3} column, which represents the first optimization configuration that outperformed -O3 (e.g., pass \emph{sroa 9} in \Cref{subfig:i5:simplifycfg34}), and on the \emph{Config.\ Removing Gains} column, which represents the configuration in which those achieved gains were lost by the addition of more optimization flags (e.g., pass \emph{simplifycfg 34} in \Cref{subfig:i5:simplifycfg34}). The second grouping appears on the \emph{Best Overall Config.} column which represents the configuration that achieved the best performance against -O3. Finally, the benchmarks within groups are sorted based on their achieved performance gains over -O3, in descending order.}
\label{fig:optsFrequencies}
\end{figure*}

The comparison of performance figures achieved after each optimizing transformation gives a direct insight into that transformation's effectiveness, relative both to preceding and subsequent optimizations, and to the ``best optimization level'' baseline. Our experiments demonstrate that for many compute kernels the best overall performance is achieved at an intermediate step of the optimization process, indicating that certain transformations applied at later optimization stages are in fact counter-productive.

The number of cases where an intermediate optimization configuration leads to a substantially better performance than the reference ``best'' optimization level -O3 is significant: 21 out of 42 benchmarks on the i5-6300U platform, and 20 out of 42 benchmarks on the Arm Cortex-A53 core achieve a performance gain of at least 3\%, and in some cases up to 71\% wrt.\ using optimization level -O3. For these benchmarks, simply stopping the optimization process at the appropriate intermediate stage provides a directly exploitable gain.

The analysis of performance degradations between consecutive transformations provides a means of improving the overall quality of the optimizations constituting the -O3 level. Each such degradation may be an artefact of a pass which opens up opportunities for subsequent optimizations. In that case the degradation should be reversed by a later transformation. However, when the degradation is not recovered later in the optimization process, it becomes a direct indication of an inadequate transformation behavior. Additionally, if the performance before degradation is better than when using the reference -O3 optimization level, it indicates a missed improvement opportunity.

\begin{table*}
    \centering
    \footnotesize
    \resizebox{\textwidth}{!}{%
    \footnotesize
      \begin{tabular}{|c|l|c|c|l|}

        \hline
\textbf{ID} & \multicolumn{1}{c|}{\textbf{Opportunity}}  & \textbf{Category} & \textbf{Target and Benchmark ID}                                   & \textbf{Location in Repository~\cite{repo:paper:results}}                     \\ \hline
1           & If-conversion heuristics                   & GI            & \begin{tabular}[c]{@{}l@{}}i5-6300u --- 8\\ ~~A53 --- 31\end{tabular} & \begin{tabular}[c]{@{}l@{}}\texttt{results/i5/benchmark-8}\\ \texttt{results/A53/benchmark-31}\end{tabular}   \\ \hline
2           & Dead code in unrolling                     & GI            & \begin{tabular}[c]{@{}l@{}}A53 --- 9\\ A53 --- 18\end{tabular}      & \begin{tabular}[c]{@{}l@{}}\texttt{results/A53/benchmark-9}\\ \texttt{results/A53/benchmark-18}\end{tabular}  \\ \hline
3           & Tuning of unrolling parameters             & TA            & \begin{tabular}[c]{@{}l@{}}A53 --- 9\\ A53 --- 18\end{tabular}      & \begin{tabular}[c]{@{}l@{}}\texttt{results/A53/benchmark-9}\\ \texttt{results/A53/benchmark-18}\end{tabular}  \\ \hline
4           & Store-vs-recompute tradeoffs               & TA            & \begin{tabular}[c]{@{}l@{}}A53 --- 10\end{tabular}              & \begin{tabular}[c]{@{}l@{}}\texttt{results/A53/benchmark-10}\end{tabular}                            \\ \hline
5           & Explicit conversion instructions           & TS            & \begin{tabular}[c]{@{}l@{}}A53 --- 7\\ A53 --- 42\end{tabular}      & \begin{tabular}[c]{@{}l@{}}\texttt{results/A53/benchmark-7}\\ \texttt{results/A53/benchmark-42}\end{tabular}  \\ \hline
6           & Better-predicted branch conditions         & TS            & \begin{tabular}[c]{@{}l@{}}A53 --- 17\\ A53 --- 36\end{tabular}     & \begin{tabular}[c]{@{}l@{}}\texttt{results/A53/benchmark-17}\\ \texttt{results/A53/benchmark-36}\end{tabular} \\ \hline
      \end{tabular} \vspace*{0.5ex}
}
\\
{\small\emph{\\ Categories: GI: General Improvement, TA: Target-Aware heuristic tuning, TS: Target-Specific optimization refinement}}
\caption{Selected compiler improvement opportunities with locations of example target code in~\cite{repo:paper:results}.}
\label{table:findings:categorized}
\end{table*}

In the following sections we illustrate one possible approach to analyzing the data produced by optimization-enhanced nightly testing. The list of findings in this illustrative study is by no means exhaustive and additional compiler improvement opportunities could be identified by further exploring the collected data. We begin the analysis with the identification of recurring sources of untapped optimization potential on the i5-6300U and Cortex-A53 platforms. We then review the reasons for the potential gains and the ways in which the potential is canceled. We mainly focus on the Cortex-A53 architecture which exhibits a more diverse range of performance and code size artefacts, and we only use the i5-6300U case for demonstrating potential cross-architecture optimization opportunities.

The findings are grouped into three categories corresponding to compiler reengineering tasks with increasing levels of knowledge and understanding of the target architectures: generic optimization improvements, target-aware heuristic tuning, and target-specific optimization refinement (\Cref{table:findings:categorized}). Generic optimization improvements are expected to be applicable to all targets, or to large classes of targets sharing a common feature such as predicated instructions or advanced branch predictors. Target-aware heuristic tuning is intended to help better exploiting the target architectures without modifying the common optimizer of a compiler. Finally, findings falling into the target-specific optimization refinement category identify the interactions between architectural mechanisms and compiler technology which cannot be easily captured in a common optimizer.

For each case discussed below, a set of supporting IR and object files is available in repository~\cite{repo:paper:results} at the location indicated in the corresponding entry of~\Cref{table:findings:categorized}. Each set contains matching IR and target object files corresponding to:
\begin{itemize}
\item the state of optimization immediately before and after the transformation that introduces the better-than-O3 performance;
\item the state of optimization immediately before and after the transformation that discards the corresponding gains;
\item the outcome of the standard -O3 optimization flow.
\end{itemize}

\subsection{Identifying recurring patterns of optimization potential}
\label{subsec:patterns_of_opts}

As shown in \Cref{fig:optsFrequencies}, there is potential for improvement over the -O3 performance baseline across recurring ranges of optimization passes. The number of benchmarks sharing a given ``opportunity range'' is a direct indication of the relevance of that range, and can be directly used to focus the compiler re-engineering effort. For each such range, the first configuration which exhibits the potential gains helps identify the unexploited feature, whereas the configuration which cancels the potential improvement 
points directly to the counter-productive optimization. Since our enhanced nightly testing system stores all IR files, the corresponding object files, and the executables for all configurations being tested, the compiler engineer can start the analysis process by reviewing the IR files generated before and after the passes that delimit each opportunity range.

\begin{figure*}[!htp]
    \centering
    \begin{subfigure}{\textwidth}
    \centering
    \begin{subfigure}{0.9\textwidth}
    \centering
      \includegraphics[trim=0.25cm 0.25cm 0.25cm 0.3cm,clip=true,width=\linewidth]{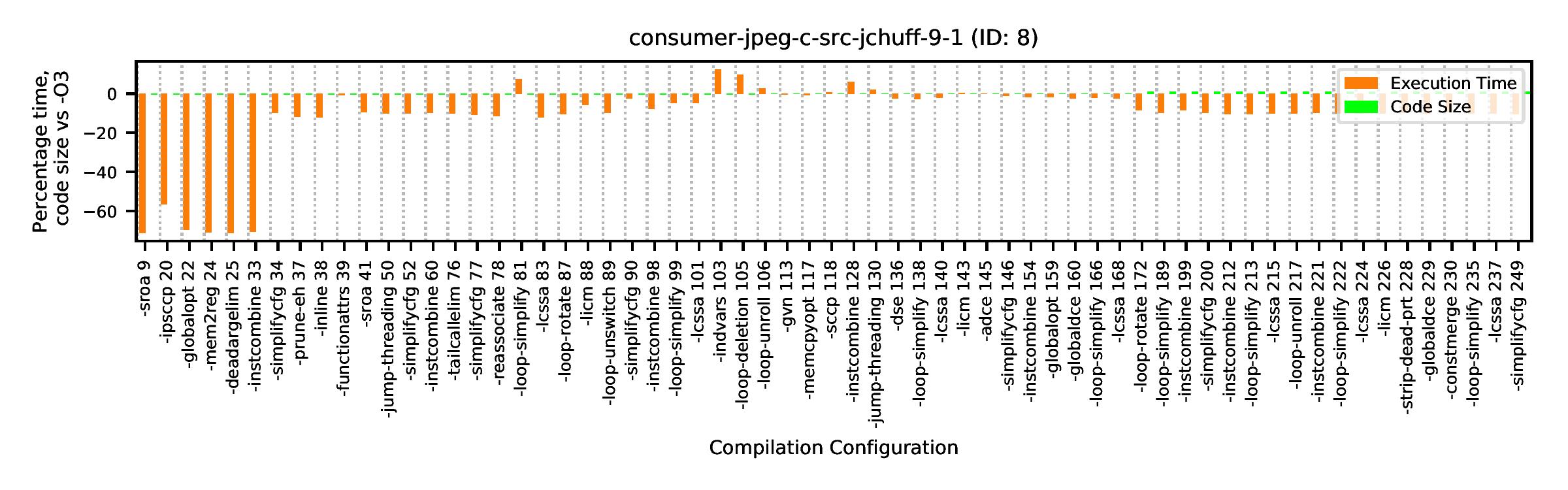}
      \subcaption{Impact of pass {\texttt simplifycfg 34} in benchmark 8 on i5-6300U}
      \label{subfig:i5:simplifycfg34}
    \end{subfigure} %\hfill
    \begin{subfigure}{0.9\textwidth}
    \centering
        \includegraphics[trim=0.25cm 0.25cm 0.25cm 0.3cm,clip=true,width=\linewidth]{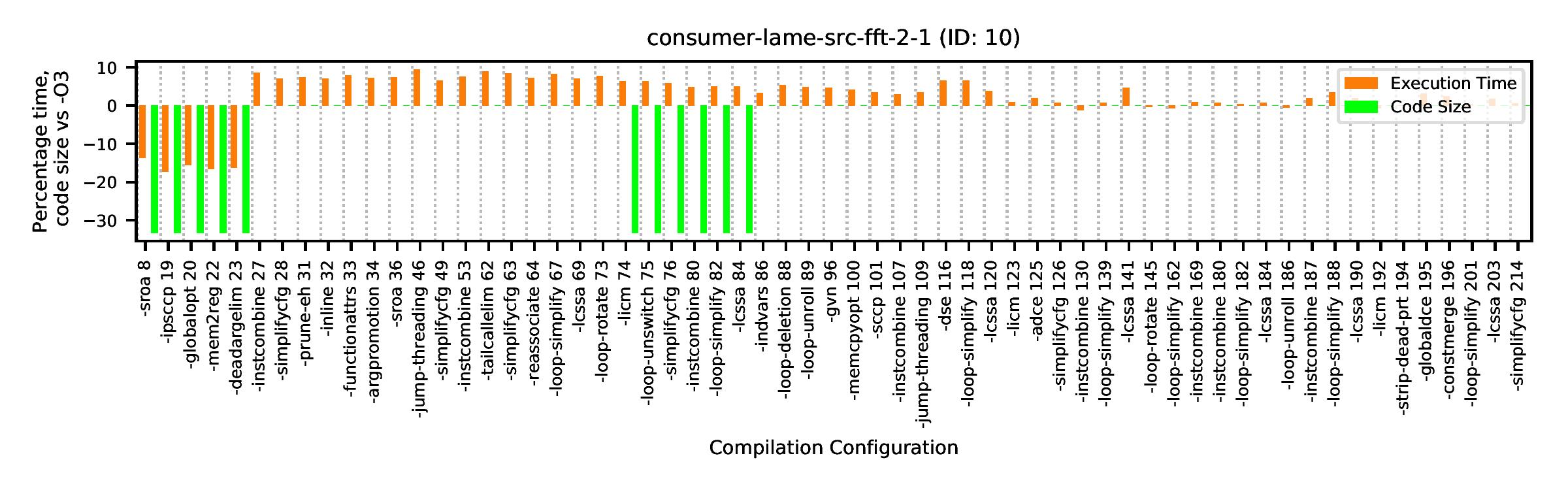}
    \end{subfigure}
      \subcaption{Impact of pass {\texttt instcombine 27} in benchmark 10 on Cortex-A53}
      \label{subfig:rpi3Bplus:instcombine27}
    \end{subfigure}
    \begin{subfigure}{\textwidth}
    \centering
    \begin{subfigure}{0.9\textwidth}
    \centering
      \includegraphics[trim=0.25cm 0.25cm 0.25cm 0.1cm,clip=true,width=\linewidth]{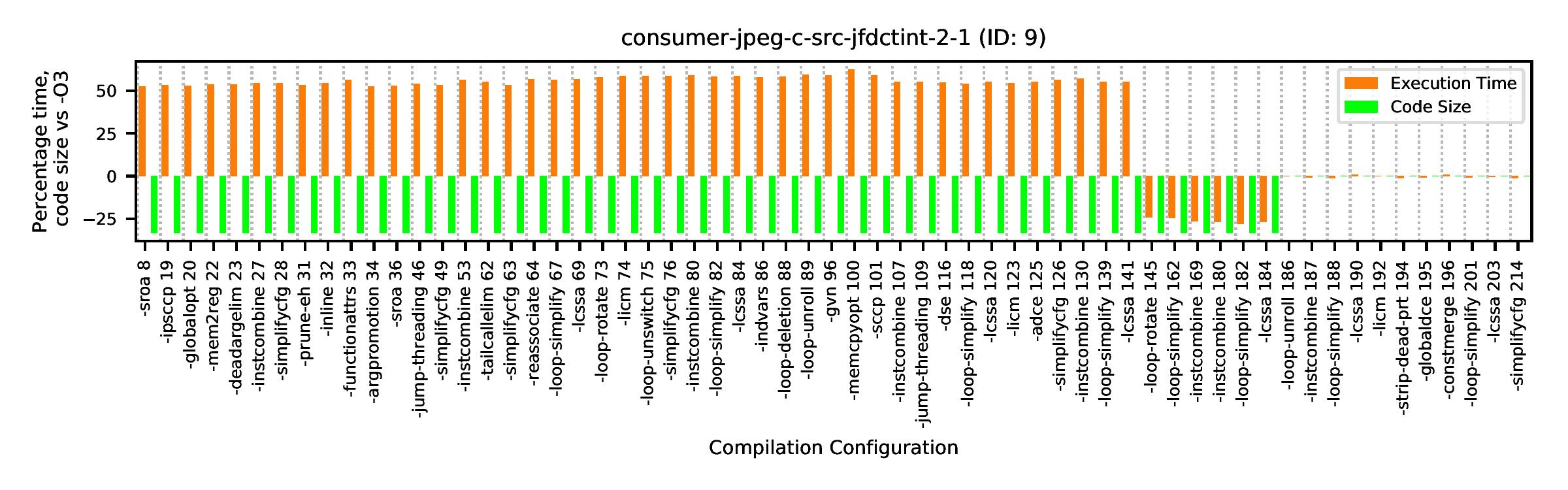}
    \end{subfigure} %\hfill
    \begin{subfigure}{0.9\textwidth}
    \centering
      \includegraphics[trim=0.25cm 0.25cm 0.25cm 0.1cm,clip=true,width=\linewidth]{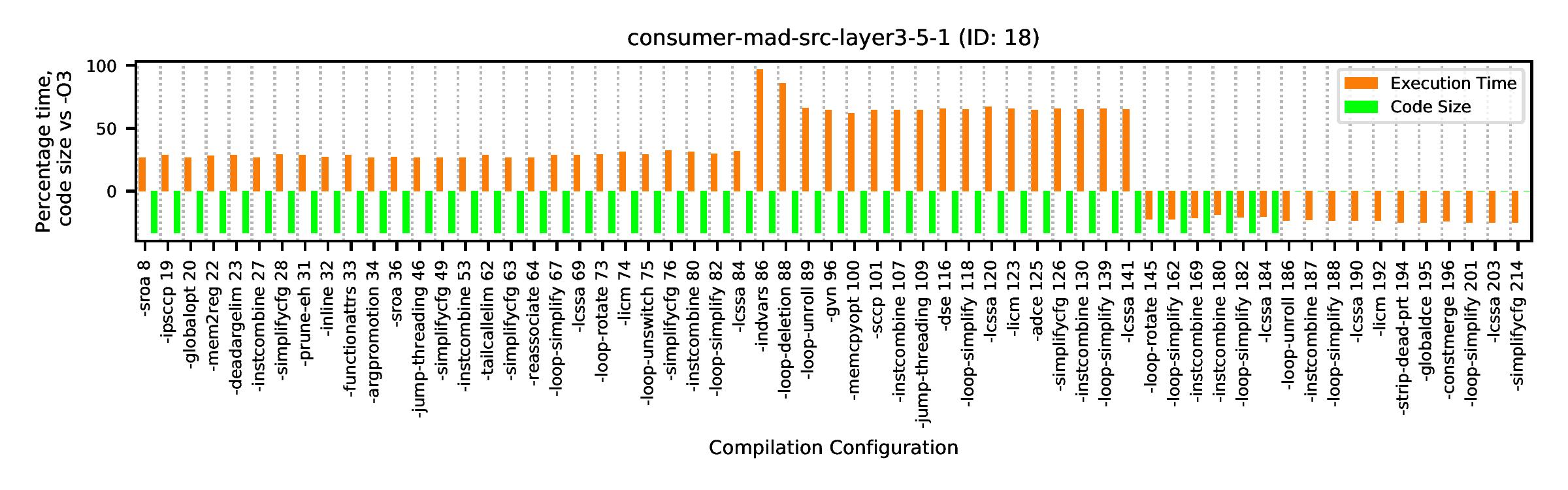}
    \end{subfigure}
    \subcaption{Interaction of passes \texttt{loop-rotate 145} and \texttt{loop-unroll 186} on Cortex-A53.}
    \label{subfig:rpi3Bplus:opt2}
    \end{subfigure}
\caption{Selected examples of better-than-O3 optimization potential. Note these figures are similar to \Cref{fig:demo_figure} but with the first 3 configurations (\emph{-O0}, \emph{-O0-custom}, \emph{simplifycfg}) removed. These configurations were significantly slower than -O3, and thus, they were obfuscating the rest of the configurations' results.}
\label{fig:rpi3Bplus:opts}
\end{figure*}

The largest cluster of optimization configurations offering hidden optimization potential on the i5-6300U architecture involves 9 benchmarks with potential performance gains of up to 71\% (cf.~\Cref{subfig:i5-regression-report}). The corresponding opportunity range begins at the first application of the static replacement of aggregates pass (\texttt{sroa 9}) and ends with the subsequent application of the control flow graph simplification pass (\texttt{simplifycfg 34}), which removes the potential gains in 10 out of 21 benchmarks.

The analysis of the detailed performance report, the IR files and the generated target code for benchmark 8 on i5-6300U (\Cref{subfig:i5:simplifycfg34}) shows that the transformation pass \texttt{simplifycfg 34} is too aggressive in applying the conversion of conditional control flow to predicated instructions (called also \emph{if-conversion}) inside the two core loops of the benchmark. The simpler of the two affected loops has the following structure:

\begin{minipage}[r]{\columnwidth}
\lstset{basicstyle=\ttfamily\footnotesize,language=C,numbers=left}
%
% ZC: This definition is needed to keep the numbering of lines inside column margins.
%     It must reside INSIDE the minipage env.
%
\makeatletter%
\def\lst@PlaceNumber{\makebox[\dimexpr 1em+\lst@numbersep][l]{\normalfont
  \lst@numberstyle{\thelstnumber}}}%
\makeatother%
%
% ZC: End of listing tuning definition.
%
\begin{lstlisting}
    c1 = -1;
    v = 1000000000L;
    for (i = 0; i <= 256; i++)
    {
      if (freq[i] && freq[i]<=v)
      {
        v = freq[i];
        c1 = i;
      }
    }
\end{lstlisting}
\end{minipage}

\begin{figure*}[!htp]
    \centering
    \begin{subfigure}[c]{0.49\textwidth}
    \centering
      \lstset{basicstyle=\ttfamily\footnotesize}
      \begin{lstlisting}
81:     mov    $0xffffffff,%r10d
87:     xor    %eax,%eax
89:     mov    $0x3b9aca00,%r11d
8f:     cmp    $0x100,%eax
94:     jbe    ab <astex_codelet__9+0xab>
96:     jmp    d0 <astex_codelet__9+0xd0>
98:     nop
...     ...
9f:     nop
a0:     add    $0x1,%rax
a4:     cmp    $0x100,%eax
a9:     ja     d0 <astex_codelet__9+0xd0>
ab:     mov    (%rdi,%rax,8),%rbx
af:     test   %rbx,%rbx
b2:     je     a0 <astex_codelet__9+0xa0>
b4:     cmp    %r11,%rbx
b7:     jg     a0 <astex_codelet__9+0xa0>
b9:     mov    %rbx,%r11
bc:     mov    %eax,%r10d
bf:     jmp    a0 <astex_codelet__9+0xa0>
...     ...
      \end{lstlisting}
      \subcaption{Optimization stopped after pass \texttt{instcombine 33}}
      \label{subfig:i5:bench8:instcombine33:code}
    \end{subfigure}
    \begin{subfigure}[c]{0.49\textwidth}
    \centering
      \lstset{basicstyle=\ttfamily\footnotesize}
      \begin{lstlisting}
81:    mov    $0xffffffff,%r15d
87:    xor    %r14d,%r14d
8a:    mov    $0x3b9aca00,%r10d
90:    cmp    $0x101,%r14d
97:    jae    c9 <astex_codelet__9+0xc9>
99:    nop
...    ...
...    ...
9f:    nop
a0:    mov    (%rdi,%r14,8),%rax
a4:    test   %rax,%rax
a7:    sete   %r11b
ab:    cmp    %r10,%rax
ae:    setg   %bl
b1:    or     %r11b,%bl
b4:    cmove  %rax,%r10
b8:    cmove  %r14d,%r15d
bc:    add    $0x1,%r14
c0:    cmp    $0x101,%r14d
c7:    jb     a0 <astex_codelet__9+0xa0>
...    ...



      \end{lstlisting}
      \subcaption{Optimization stopped after pass \texttt{simplifycfg 34}}
      \label{subfig:i5:bench8:simplifycfg34:code}
    \end{subfigure}
\caption{Machine code impact of pass \texttt{simplifycfg 34} on the first innermost loop of benchmark 8 on i5-6300U.}
\label{fig:i5:bench8:machinecode}
\end{figure*}
At each iteration, the assignment statements in source lines 7 and 8 above are only executed if several conditions are met at once (source line 5); if not, program execution advances to the next iteration.
The machine code obtained after stopping the IR optimization process at pass \texttt{instcombine 33}~(\Cref{subfig:i5:bench8:instcombine33:code}) is characterized by the presence of early conditional branches to the \emph{loop latch} block at address \texttt{0xa0} and which handles the termination of the current iteration. These early branches shorten the actual critical path of many iterations resulting in a very fast execution of the benchmark. \Cref{subfig:i5:bench8:simplifycfg34:code} shows the assembly code generated after pass \texttt{simplifycfg 34} has also been applied. 
% However, 
The application of the \texttt{simplifycfg 34} pass removes the early conditional branches and produces a single basic block with a unique conditional branch placed after the loop latch code which now starts at address \texttt{0xbc}~(\Cref{subfig:i5:bench8:simplifycfg34:code}). The assignments from source lines 7 and 8 are now implemented using predicated assignment instructions at addresses \texttt{0xb4} and \texttt{0xb8}; the branch condition is computed using predicated assignments at addresses \texttt{0xa7} and \texttt{0xae} and a Boolean instruction at address \texttt{0xb1}. The computation of the loopback condition introduces a very long critical path which is taken in all iterations. As a result, the average execution time of the complete benchmark increases three-fold, thus cancelling almost the entire gain potential available upon entering pass \texttt{simplifycfg 34}. Clearly, this behavior calls for a careful revision of if-conversion strategies in the compiler and is a natural opportunity for a \emph{generic optimization improvement} that could benefit multiple targets, categorized as case~1 in \Cref{table:findings:categorized}. Furthermore, in~\Cref{subsec:caseStudy} below we provide an example case study which illustrates the improvements achievable with a simple modification of the \texttt{simplifycfg} transformation pass discussed above.

The largest cluster of similarly behaving benchmarks on the Cortex-A53 (see \Cref{subfig:rpi3Bplus-regression-report}) consists of benchmarks 10, 36, 42, 31, and 7. Its corresponding opportunity range starts with the first ``static replacement of aggregates'' (\texttt{sroa 8}) pass and ends with the first application of the instruction combiner (\texttt{instcombine 27}) pass. The instruction combiner pass removes many of the explicit conversion instructions and performs selective \emph{if-conversion}. A deeper analysis of the IR files and the generated target code for the benchmarks of the cluster leads to a broad range of findings:

\begin{itemize}
\item In benchmark 31, the source of the hidden performance potential is the presence of explicit conditional control flow with unbalanced workloads in the ``true'' and ``false'' paths. The instruction combiner pass replaces the explicit conditional control flow structure with predicated instructions, thus aligning the critical path of the resulting code on the longest of the critical paths of the original control flow structure. Like in the case of i5-6300U and the \texttt{simplifycfg 34} pass, this issue signals a deficiency of the \emph{if-conversion} strategy and is an example of a \emph{general optimization improvement} which can benefit all targets. The similarity with the case of benchmark 8 on i5-6300u suggests that the two deficiencies of \emph{if-conversion} may have to be addressed in conjunction, and have therefore been grouped together as case~1 in \Cref{table:findings:categorized}.
\item In benchmark 10 (\Cref{subfig:rpi3Bplus:instcombine27}, case~4 in~\Cref{table:findings:categorized}), the presence of an explicit conversion instruction forces the recomputation of a value which would otherwise require an additional register. The corresponding reduction in register pressure increases the performance and reduces both memory traffic and the actual code size. This case can lead to \emph{target-aware heuristic tuning} of store-vs.-recompute tradeoffs.
\item The presence of explicit conversion instructions enables the recognition of complex instruction patterns involving explicit conversions (multiply-accumulate in benchmark 42 and addition/subtraction with operand shift in benchmark 7, case~5 in \Cref{table:findings:categorized}) and the use of seemingly faster branch instructions (conditional branches on signed rather than unsigned comparison conditions in benchmark 36, case~6 in \Cref{table:findings:categorized}). These three cases are linked to the specific instruction set and the microarchitectural behavior of the target architecture and belong to the category of \emph{target-specific optimization refinements}.
\end{itemize}

The opportunity ranges opened on Cortex-A53 by loop rotate passes (\texttt{loop-rotate 73} and \texttt{loop-rotate 145}) are associated with loop transformations. Optimization opportunities offered by the second loop rotation pass (\texttt{loop-rotate 145}, cf.~\Cref{subfig:rpi3Bplus:opt2}) are more significant and illustrate a changing behavior of the compiler regarding the interactions between loop vectorization and loop unrolling.

In benchmark 9 (upper graph of \Cref{subfig:rpi3Bplus:opt2}), pass \texttt{loop-rotate 145} vectorizes the original loop, yielding an outer loop with only two iterations and a performance improvement of 27.7\% over the code generated using the standard -O3 optimizations. The subsequent unrolling of the outer loop in pass \texttt{loop-unroll 186} fully unrolls the loop body producing code that is fully sequential but twice as large, and the corresponding performance loss may be caused by instruction cache thrashing artefacts.

In contrast, in benchmark 18 (lower graph of~\Cref{subfig:rpi3Bplus:opt2}) a similar performance gain is achieved through loop vectorization, but it is not canceled by a subsequent loop unrolling of the vectorized loop. This difference in behavior is explained by the fact that quantitative settings of the loop-unroll pass depend on the optimization flag used when invoking the optimizer. Our optimization sequences start from level -O0 and therefore, the loop unrolling passes use the default loop unrolling threshold value applied at optimization levels lower than -O3.

On the other hand, the standard optimization sequence of the -O3 level uses a default unroll threshold value which is twice as large, enabling the unrolling where our partial optimization sequences prevent it. This artefact raises the importance of \emph{target-aware heuristic tuning} (case~3 in \Cref{table:findings:categorized}) which requires significant understanding of the target architecture, but may be needed to utilize the target architecture at its best. It also demonstrates that the compiler should actively leverage the information about the current target architecture instead of relying on heuristically determined target-independent threshold values.

In \Cref{subfig:rpi3Bplus:opt2}, the final loop unrolling pass (\texttt{loop-unroll 186}) not only impacts performance, but also cancels the potential for code size reduction observed in benchmarks 9 and 18. The increase in code size caused by this optimization pass is linked to the introduction of additional ``catch-up'' loops intended to handle the cases where the actual number of iterations is not known beforehand and might not be a multiple of the unrolling factor. However, in the tested benchmarks the loop has a constant number of iterations and once vectorized, it is fully unrolled to linear code. This means that the catch-up loops are redundant and should be removed, yet they are actually left in the code calling for a \emph{generic optimization improvement} (cf.~case~2 in~\Cref{table:findings:categorized}).

As a last example, the analysis of the behavior of benchmark 17 on Cortex-A53 leads to a potential \emph{target-specific optimization refinement} (case~6 in \Cref{table:findings:categorized}): the loss of performance potential during the first loop-rotate pass (\texttt{loop-rotate 73}) corresponds to the inversion of conditional branch conditions in the benchmark core loop, with all other instructions of the core loop remaining identical. The associated 16.2\% decrease in code performance hints at a branch prediction artefact that could be related to the findings of benchmark 36 (described above) in which the benchmark performance is directly linked to the relative execution times of signed vs.\ unsigned conditional branch instructions.

\subsection{Leveraging the identified optimization opportunities}

The example findings described in the preceding section suggest that our approach of testing the quality of partial optimization configurations in compilers can benefit the compiler technology community, industrial users and developers of compilers, as well as hardware architects. Ideally, our approach should be integrated into the compiler development flows so that the findings can be systematically collected, reviewed and dispatched according to their scope. \emph{Generic optimization improvement} opportunities, once identified, should be reported to compiler maintainers and the compiler technology community at large. The resulting improvements in the given compiler will benefit all developers and users of that compiler on a wide range of platforms.

\emph{Target-aware heuristic tuning} opportunities are of particular importance to developers and maintainers of industrial compilers who focus on the best possible utilization of their target architectures. The findings can help selecting the most appropriate values of quantitative parameters of transformations, if these parameters can be controlled by the user (such as the loop-unrolling threshold), and can identify the cases where new parameters should be introduced.

The performance potential identified in nightly routine tests can then be easily made available to users, e.g.\ by supplying sets of parameter options tuned for the different configurations of the target architecture. In addition, the performance of the generated code can be finely matched to the target platforms without affecting the basic principle of a common optimizer and without having to modify the optimizer code (with all the quality risks this would imply.)

Finally, \emph{target-specific optimization refinements} identify subtle interactions between architectural mechanisms and compiler technology which may require a coordinated effort of the hardware and compiler communities. This category of findings requires by far the deepest levels of hardware architecture and compiler technology knowledge. Findings regarding the behavior of branch predictors, for example, can simultaneously provide useful feedback to hardware architects and to compiler developers. The former can gain additional awareness of the ways the branch prediction is behaving on compiler-generated code, and the latter will be able to review the flow of predictor-aware code generation. We have seen in the previous section that branch prediction and code generation may interfere in significant ways.

In order to assess the actual impact of these interactions, the static analysis of generated code may prove insufficient, requiring detailed information about the behavior of specific micro-architectural features of the target platform, e.g.\ in the form of data from hardware performance counters \cite{Oprofile:online}. The use of performance counters requires a good understanding of the target architecture, making them a tool aimed primarily at expert compiler engineers.

However, once the correlation between specific hardware events, the readings of the performance counters and the effects of a given optimization has been established, the monitoring of the relevant hardware events can be integrated into the nightly routine tests as an additional metric to be tracked in addition to execution time, code size or energy consumption.

\subsection{Case Study: Improving the SimplifyCFG Pass of LLVM}
\label{subsec:caseStudy}

Based on the findings of~\Cref{subsec:patterns_of_opts} we modified the implementation of pass \texttt{simplifycfg} of LLVM 6.0.1 release compiler to better leverage loop-related information. The operation of the \texttt{simplifycfg} pass was modified so that whenever the basic information about loops is known, the pass skips the if-conversion of conditionals which may result in an early termination of the current loop iteration. This ensures that the fast execution path is not merged with slower paths containing speculatively executed instructions, making it possible for branch predictors to exploit the short path much more efficiently. The corresponding patch is available in repository~\cite{repo:paper:results} under path \texttt{code/llvmorg-6.0.1-simplifycfg-opt.patch}.

We evaluated this modification on both the i5 and the Cortex-A53 platforms.  \Cref{tab:implemented_opt} summarizes the results for the six benchmarks which on the i5 platform exhibited the largest performance degradations associated with the application of pass \texttt{simplifycfg} as reported in~\Cref{subfig:i5-regression-report}. The other benchmarks were not affected by this modification in any significant manner.  For benchmark 8 which was the most severely affected one on the i5 platform, the modification unlocked the entire latent improvement potential leading to a 2.5-fold performance increase.  On Cortex-A53 the modification also led to a major improvement, in excess of the potential identified in the nightly testing campaign. On both platforms, the changed behavior of pass \texttt{simplify 34} enabled a more effective code generation in the compiler backend. In benchmark 24, the i5 performance improvement potential remains now unaffected by the successive applications of the modified \texttt{simplifycfg} pass. However, the potential two-fold increase in performance available after pass \texttt{simplify 34} is now cancelled by the application of pass \texttt{loop-rotate 87}, indicating that this specific pass should become the subject of a new in-depth investigation. Finally, the modification of the \texttt{simplifycfg} pass appears to unlock additional improvement potential in the initial application of the pass at the beginning of the optimization process (\texttt{simplifycfg 7}). While in benchmark 35 the relative reduction in overhead observed at pass \texttt{simplify 34} is not reflected in the performance at optimization level -O3, in benchmarks 37 and 42 our modification leads to a 2.1\% and 5.4\% improvement over -O3 performance, respectively.

% \usepackage{multirow}
% \usepackage{graphicx}
\begin{table*}
\centering
\resizebox{\textwidth}{!}{%
\footnotesize
\begin{tabular}{|c|r|r|r|r|r|r|}
\hline
\multicolumn{1}{|c|}{\multirow{3}{*}{\textbf{Benchmark ID}}} & \multicolumn{3}{c|}{\textbf{Intel i5-6300U}} & \multicolumn{3}{c|}{\textbf{Arm Cortex-A53}} \\ \cline{2-7}
\multicolumn{1}{|c|}{} & \multicolumn{2}{c|}{\textbf{After pass simplifycfg34}} & \multicolumn{1}{c|}{\textbf{-O3}} & \multicolumn{2}{c|}{\textbf{After pass simplifycfg34}} & \multicolumn{1}{c|}{\textbf{-O3}} \\ \cline{2-7}
\multicolumn{1}{|c|}{} & \multicolumn{1}{c|}{\textbf{Without patch}} & \multicolumn{1}{c|}{\textbf{With patch}} & \multicolumn{1}{c|}{\textbf{With patch}} & \multicolumn{1}{c|}{\textbf{Without patch}} & \multicolumn{1}{c|}{\textbf{With patch}} & \multicolumn{1}{c|}{\textbf{With patch}}  \\ \hline
2 & -2.33\% & -2.33\% & -2.33\% & -2.15\% & -2.15\% & 0.00\% \\ \hline
8 & -11.53\% & -60.51\% & -61.36\% & -30.99\% & -45.72\% & -36.07\% \\ \hline
24 & -47.83\% & -50.00\% & -2.17\% & 1.39\% & 1.29\% & 0.87\% \\ \hline
35 & 64.29\% & 57.14\% & 0.00\% & 15.75\% & 15.77\% & 0.10\% \\ \hline
37 & 11.15\% & 7.69\% & -2.12\% & -2.11\% & -2.30\% & 0.04\% \\ \hline
42 & 14.86\% & 12.16\% & -5.41\%  & 8.77\% & 9.17\% & -0.45\% \\ \hline
\end{tabular}%
}
\caption{Relative change of performance of code compiled using LLVM 6.0.1 with and without the proposed \texttt{simplifycfg} patch applied to the optimization configuration stopping after the \emph{simplifycfg34} pass, and when using -O3 with the proposed patch (baseline used is the performance of code compiled using the unpatched optimizer and the -O3 optimization level).  Negative values indicate improvement, and positive values indicate degradation.}
\label{tab:implemented_opt}
\end{table*}

\section{Related Work}
\label{sec:related_work}

Auto-tuning of compiler optimizations has emerged in the last decade, taking two main forms; iterative and machine-learning-based (MLB) compilation~\cite{Boyle:2018,Ashouri:survey}. Typically, the aim is to find new optimization sequences that can outperform what the standard compiler optimization levels can achieve in terms of effective resource usage for a particular program and architecture; the resource of interest being execution time, energy consumption or memory usage (code size). 
The motivation for automatic tuning is that the possible optimization configuration space is too large to be explored in practice, and thus, hidden optimization opportunities can exist within that space. 
%These can outperform the standard optimization levels for a specific architecture or programming language. 
For example, GCC v4.7 has $2^{82}$ possible optimization combinations~\cite{Pallister2015}, not counting the possible values of quantitative parameters. The concept of common architecture-independent optimizers, while helping compiler developers in supporting more programming languages and more architectures, has the adverse effect of preventing high-level optimizations from matching target architectures' quantitative characteristics. This can produce suboptimal executables in terms of efficiently using a specific architecture's resources.

Iterative compilation typically randomly samples the optimization configuration space until finding a configuration that outperforms a predefined optimization level~\cite{Ashouri:2017}. The technique has in many cases proven to provide significant performance gains~\cite{bodin:inria-00475919,CK:2018}, but typically a large number of optimization configurations, in the order of hundreds to thousands, need to be evaluated before reaching any performance gains over standard optimization levels. Thus, iterative compilation has been traditionally used as a baseline to assess the performance of MLB compiler auto-tuning techniques~\cite{Fursin2011,Ashouri:2017,blackmore2015}. 
MLB techniques aim to beat the performance of iterative compilation by finding a better optimization configuration in a shorter time. Thus, MLB techniques try to strategically sample the optimization configuration space based on the models built during their training phase. These models are being trained on either static code features~\cite{Fursin2011} or profiling information~\cite{Cavazos:2007}, such as performance counter values that characterize the programs in the training set, and a performance metric for the dependent variable. An example of such a performance metric is the execution time of programs when applying a specific optimization configuration.

Typically, these techniques require a large training phase~\cite{Ogilvie:2017} to create the predictive models they rely on. Furthermore, they are hardly portable across different compilers, different versions of the same compiler, or different architectures. Even if a single flag is added to the set of a compiler's existing flags, the whole training phase has to be repeated. Moreover, extracting some of the metrics that these techniques depend on, such as static code features, might require a significant amount of engineering~\cite{Boyle:2018}. Thus, MLB techniques are inadequate for systematic testing and improvement of compilers.

While iterative compilation and MLB approaches can assist software developers in improving an application's resource usage by auto-tuning the compiler settings, they offer limited value to the compiler engineer on how to improve the compiler's common optimizer. This is because they typically offer limited information in regards to the potential causes of their achieved gains over a standard optimization level. 
Moreover, MLB approaches only provide predictions of good optimization sequences based on an application's features. These might or might not work well for applications unseen by the machine-learning training phase. Compiler engineers need more concrete evidence to guide their efforts of tuning the compiler's common optimizer.

Our enhanced nightly testing system, introduced in \Cref{sec:exposeOpts}, offers a different approach which can assist the compiler engineer to ``debug'' the compiler optimization sequences in terms of their effectiveness in a systematic way. This is due to: 
a) the ability of our technique to attribute the optimization effects observed to specific transformation passes exercised in an optimization sequence, and 
b) the technique offering concrete data to drive the tuning of the common optimizers, instead of MLB predictions. 

Although a number of the existing approaches are able to provide some insights to the compiler engineer in regards to the effectiveness of optimization configurations, they were not designed to support this as part of the daily development cycle of compilers. For example, in \cite{Dubach:2009}, the authors devise a MLB approach which can account for micro-architecture features and performance monitoring counters, and thus, the technique is portable across micro-architectures. Although, the technique was able to quantify the effectiveness of optimization flags through a posthoc analysis, porting the approach to a new version of the compiler would require retraining their expensive machine learning model, seven million training points plus cross-validation. 

The Combined Elimination technique introduced in~\cite{Zhelong:2006} was shown able of exposing optimization tuning opportunities with an acceptable overhead for a nightly testing system, i.e.\ within the range of hours for the tested architectures. The technique deploys heuristics to disable flags of the -O3 standard optimization level of the GCC 3.3.3 compiler, to find the optimization flags that degrade performance. Although it can be argued that the technique exposes optimization patterns and their sources, disabling optimization flags was performed without validating that the compiler was indeed behaving as instructed. Both LLVM and GCC compilers have mechanisms to automatically enable optimization and analysis passes when an optimization configuration does not meet all the required dependencies. Thus, the outcome of disabling a specific optimization might be related to other side-effects of the compiler trying to recover broken dependencies that were caused by the eliminated optimization. Furthermore, the paper mentions no validation of the correctness of the generated executables.

Energy consumption of computing is becoming critically important for economic, environmental, and reliability reasons~\cite{Eder:2016:ENTRA,Georgiou2017}. In~\cite{Georgiou:2018:SCOPES}, the technique also used in this paper for exploring the standard optimization levels, was able to accurately account for energy consumption through physical hardware measurements on deeply embedded devices.~In future work, we will explore if energy profilers~\cite{RAPL:online} can achieve the same for platforms with higher-end architectures that do not allow for processor's direct energy measurements, such the ones explored in this paper.

\section{Conclusion and Future Work}
\label{sec:conc_future}

In this paper, we first define the four properties a technique should exhibit to support the tuning of a compiler's optimizer as part of the compiler's daily development cycle: \emph{portability}, \emph{agility}, \emph{versatility}, \emph{insightfulness}. The majority of the traditional auto-tuning techniques, such as iterative compilation and MLB approaches, are mainly engineered to provide better optimizations than the standard optimization levels for a specific application, rather than to support the compiler engineer in tuning a standard optimization level. 
These techniques tend to require a new expensive training phase at each compiler update, or need to run for thousands of iterations. Furthermore, with few exceptions, such as the combined elimination technique \cite{Zhelong:2006}, they fail to expose optimization patterns and their possible causes. Thus, they fail to exhibit all four required properties, and in practice, they are of limited value to the compiler engineer.

We, therefore, propose a new take on the classic nightly testing system of compilers, which exhibits all four required properties. The system is enhanced with statistics that can systematically expose the behavior of the standard compiler optimization levels wrt.\ performance. To achieve this, we adopt the technique proposed in \cite{Georgiou:2018:SCOPES}, i.e.\ we exploit prefix subsequences of the standard optimization levels rather than arbitrary permutations of optimizations. Thus, in contrast with iterative compilation or MLB techniques, our approach offers compiler engineers an intuitive way of correlating performance variations with the internal structure of the optimizer. To the best of our knowledge, our approach is the first that focuses on providing a systematic means of tuning the compiler's standard optimization levels and can be easily integrated into the daily development cycle of the compiler.

By applying the technique to benchmarks from the LLVM test-suite, we established the existence of significant optimization opportunities within the standard optimization levels, firstly, on more complex architectures (the x86-64 based i5-6300U and the Armv8-A based Cortex-A53) than the deeply embedded ones used in~\cite{Georgiou:2018:SCOPES}, and secondly, across multiple versions of the LLVM compiler, namely the LLVM v3.8 and v5.0 examined in~\cite{Georgiou:2018:SCOPES} and also the v6.0.1 examined in this paper. Moreover, the short time required for porting the technique to new architecture and compiler's versions, within less than an hour, ensures the required \emph{portability} property. 

Significant performance gains were observed for more than half of the 42 benchmarks tested, with an average of 11.5\% and 5.1\% execution time improvement for the i5-6300U and the Cortex-A53 processors, respectively. These results were collected, classified and exploited by the proposed nightly testing system to expose a series of potential architecture-depended and cross-architecture optimization patterns and rank them according to their potential impact; fulfilling the \emph{versatility} required property. Furthermore, the small number of optimization configurations that needs to be exploited, defined in~\Cref{sec:comp_and_analysis}, ensures minimal overhead on the nightly testing system, fulfilling the \emph{agility} property.

This is of significant value for compiler engineers who can focus their efforts on exploiting the hidden gains and removing the shortcomings of the key performance-affecting optimizations. The resulting insights may lead to cross-architecture optimizer improvements that benefit all users of the compiler, to architecture-specific tuning relevant for suppliers and users of industrial compilers, and to new ways of handling innovative hardware mechanisms at compiler level. To the best of our knowledge, this is the first work on automated tuning of compilers that enables the discovery and the analysis of new optimization potential to this extent. 
In a case study of optimization refinement, we leveraged the insightfulness of our approach to identify and remove a significant shortcoming of the CFG simplification pass of the LLVM v6.0.1 compiler, allowing the compiler's optimizer to achieve the observed performance gains over the optimization level -O3 for the majority of the affected benchmarks.

We expect that our technique can be applied to any compiler framework which supports command-line activation/deactivation of individual compiler passes. We intend to apply it to the target-independent optimization passes of GCC in order to evaluate the latent performance potential of GCC and the ability of the technique to support the improvement of more mature, yet harder-to-maintain compilers. This will also further validate our intra-compiler portability claim wrt.\ different versions of GCC. Beyond that, we expect to gain deeper understanding of the cross-compiler applicability and portability of our technique. 

In the future, we plan to extend our nightly testing system with the collection of hardware performance counter data to further support compiler engineers in identifying and exploiting potential optimization opportunities that are not statically analyzable and may be linked to micro-architectural features of the target processors. 

Another research direction we are currently exploring is to extend the prefixes that perform better than the standard optimization levels by attaching to them beneficial optimization passes selected by either machine learning or iterative compilation. Such an approach will allow us to potentially discover even better optimization sequences than we currently do, but with lower exploitation overheads than the ones of machine learning and iterative compilation techniques that start their exploitation from scratch.

Future work can also focus on tuning the compiler's standard optimization level in regard to other resources such as energy consumption and code size. As shown in \Cref{fig:all_results_graphs} and in~\cite{Georgiou:2018:SCOPES}, our approach can offer improvements over the code size offered by the standard optimization levels. This can be exploited in a more relevant context; for example, the tuning of the -Oz optimization flag for small embedded processors, such as the Arm Cortex-M series.

Finally, our technique of tuning the performance of the standard optimization levels should become standard practice for compiler development. Thus, we plan to enhance the existing LLVM test suite framework with our technique through an LLVM plug-in. This will have an immediate practical benefit to compiler engineers.

\section*{Acknowledgments}
This research is supported by the European-Union's Horizon 2020 Research and Innovation Programme under grant agreement No. 779882, TeamPlay (Time, Energy and security Analysis for Multi/Many-core heterogeneous PLAtforms).

\bibliographystyle{alphaurl}
\bibliography{typeinst}

\end{document}